\documentclass[twocolumn,showpacs,showkeys,preprintnumbers,superscriptaddress,amsmath,floatfix,amssymb,secnumarabic,nofootinbib]{revtex4-1}
\usepackage[colorlinks=true]{hyperref}
\usepackage{graphicx}
\usepackage{dblfloatfix}    
\usepackage{placeins}
\usepackage{braket}
\usepackage{amsmath}
\usepackage{physics}
\usepackage{epsfig}
\usepackage{float}
\usepackage{nicefrac}
\usepackage{xcolor}
\usepackage{color}
\usepackage{bm}
\usepackage{subfigure}
\usepackage{chngcntr}
\usepackage[normalem]{ulem}

\def\({\left(}
\def\){\right)}
\def\[{\left[}
\def\]{\right]}

\newcommand{\eV}{\, {\rm eV}}

\newcommand{\ve}[1]{\boldsymbol{#1}}

\newcommand{\beq} {\begin{eqnarray}}
\newcommand{\eeq} {\end{eqnarray}}

\begin{document}
\sloppy

\title{ Hydrodynamics of particle-hole symmetric systems: a quantum Monte Carlo study}

\author{Adrien~Reingruber}
\email{adrien.reingruber@uni-wuerzburg.de}
\affiliation{Institut f\"ur Theoretische Physik und Astrophysik, Universit\"at W\"urzburg, 97074 W\"urzburg, Germany}

\author{Kitinan~Pongsangangan}
\email{kitinan.pon@mahidol.ac.th}
\affiliation{Department of Physics, Faculty of Science, Mahidol University, Bangkok 10400, Thailand}
\affiliation{Institute of Theoretical Physics, Technische Universit\"at Dresden, 01062 Dresden, Germany}

\author{Fakher~Assaad}
\email{fakher.assaad@uni-wuerzburg.de}
\affiliation{Institut f\"ur Theoretische Physik und Astrophysik, Universit\"at W\"urzburg, 97074 W\"urzburg, Germany}
\affiliation{W\"urzburg-Dresden Cluster of Excellence ct.qmat, Am Hubland, 97074 W\"urzburg, Germany}

\author{Maksim~Ulybyshev}
\email{maksim.ulybyshev@uni-wuerzburg.de}
\affiliation{Institut f\"ur Theoretische Physik und Astrophysik, Universit\"at W\"urzburg, 97074 W\"urzburg, Germany}

\begin{abstract}
The emergence of hydrodynamic behavior in electronic flow within clean, particle-hole-symmetric systems at half-filling is a non-trivial problem. Navier-Stokes (NS) equations describe the momentum flow, while experimental measurements typically capture the current flow profiles. However, in particle-hole-symmetric systems, electric current and momentum flow are entirely decoupled because electrons and holes move in opposite directions with equal distribution functions. This makes it challenging to link NS equations to observed flow patterns. In this work, we  demonstrate that the hydrodynamic behavior of the charge current at half filling can emerge despite the absence of momentum flow.  By combining Boltzmann transport theory with numerically exact Quantum Monte Carlo simulations of clean graphene samples, we show that NS-type equations can be derived directly for the charge current, eliminating the need for any additional mechanism coupling the velocity field and charge current in explaining the experimentally observed hydrodynamic flow profiles in graphene at half-filling. We show that a new transport quantity  - the current diffusion coefficient - replaces viscosity and expect this description to be valid for any particle-hole symmetric system. Our results provide new insights into the interpretation of experimental data and demonstrate how Quantum Monte Carlo calculations can serve as an alternative to experiments in transport measurements to verify the kinetic theory results. 
\end{abstract}

\maketitle

{\it Introduction.}

Strongly correlated systems provide an exciting playground to study the electronic transport due to the presence of multiple scattering mechanisms generating different time scales. In particular, when the particle-particle relaxation time, $\tau_{ee}$, is much shorter than the diffusive scattering time, $\tau_{\text{diff}}$ (arising from factors such as defects), the system transitions from Ohmic electron transport to a regime resembling the hydrodynamics of a viscous fluid \cite{Gurzhi68,deJong95,PhysRevB.56.8714,Fritz2009,Lucas_2018,Ho18}. This transition significantly alters the current flow profiles, with complex patterns like vorticity appearing depending on the sample geometry\cite{Levitov2016}. The dominance of particle-particle scattering parallels the behavior of strongly correlated quantum field theories (QFTs)\cite{KARSCH2008217, PhysRevLett.94.111601}, and it has been proposed that in graphene at the charge neutrality point (CNP), we may approach the lower bound for the ratio of shear viscosity to entropy density \cite{Fritz2009}. A limit initially derived for a broad class of strongly correlated QFTs \cite{PhysRevLett.94.111601}.

Recent advances in experimental techniques have provided an additional boost to these studies. Within now exceptionally clean graphene samples encapsulated between two layers of boron nitride, particle-particle collisions become the dominant scattering mechanism, thereby shifting electronic transport to the hydrodynamic regime. Combined with new imaging techniques, this development enables direct visualization of hydrodynamic flow profiles and measurement of viscosity in graphene \cite{Bandurin2018, Sulpizio2019,Ku2020,PhysRevLett.129.087701} even at CNP.

 However, there is an apparent contradiction in the interpretation of these experiments at CNP, an issue raised already in \cite{Narozhny_2021}. Viscosity appears in the Navier-Stokes (NS) equations for the velocity field, and they essentially describe the dynamics of momentum density under external forces. On the other hand, the momentum flow itself is usually induced in experiments by applying an external electric field to the sample. Thus, the validity of the hydrodynamic description relies on the response of the momentum $\hat{P}^\alpha$ to the perturbation caused by the coupling of the electric field to the charge current $\hat{j}^\gamma$.  Within Kubo linear response the momentum induced by the electric field is evaluated from the correlation function $\operatorname{Tr}\left( e^{-\beta \hat H}  \left[\hat P^\alpha(t), \hat j^\gamma(0) \right] \right)$. Since $\hat{P}^{\alpha}$ ($\hat j^\gamma$)  is even (odd) under particle-hole symmetry the correlator vanishes for any  particle-hole symmetric model system. (see Supplementary Material (SM) \cite{Supplement}).
 
Physically, this corresponds to electrons and holes with equal distribution functions propagating in opposite directions. Away from CNP, this is not a problem, as electrons and holes do not fully compensate each other, and the response of momentum density to the electric field is nonzero, leading to the standard hydrodynamic description of the flow. However, at CNP, the momentum density governed by the NS equations remains unaffected by an external electric field and averages to zero \cite{Levitov2016}.

Now the question arises, why do we still see the hydrodynamic behavior of the charge current in experiments in graphene at CNP, since graphene with its emergent Lorentz symmetry is just one example of a system with particle-hole symmetry.

In this paper, we demonstrate that the hydrodynamic behavior of the charge current at CNP can emerge despite the absence of momentum flow. Typical imaging techniques \cite{Ku2020,PhysRevLett.129.087701} visualize the stray magnetic field, capturing the charge current flow rather than the momentum distribution. Previously, these experiments were still interpreted in terms of momentum flow and shear viscosity, computed in \cite{Fritz2009}, assuming some connection between momentum and current even at CNP (e.g. via an artificial chemical potential $\mu = k_B T$ \cite{Ku2020}). Here we develop the hydrodynamic description directly for the charge current, completely circumventing momentum. For the first time, we obtain NS-type equations for the current density, and a new transport coefficient, the current diffusion coefficient $\zeta$, replaces viscosity. 

To show the emergence of hydrodynamics of the charge current, we combine numerically exact Quantum Monte Carlo simulations of a microscopic Hamiltonian with Boltzmann transport theory. Due to absence of the sign problem, QMC provides exact results \cite{Sorella_1992, Troyer05, Ulybyshev13, Hohenadler14, Jiménez2021}. This allows us to observe hydrodynamic behavior of charge current in a well-controlled environment, where all scattering mechanisms are fixed by the parameters of the microscopic Hamiltonian. For example, we exclude phonons and restrict diffusive scattering to the sample's edges, leaving the bulk completely clean, with particle-particle collisions as the sole scattering mechanism. After obtaining the current profiles from QMC, we extract the transport coefficients by fitting the profiles to solutions of the corresponding current flow equations derived from the Boltzmann transport equation. Qualitative agreement between QMC and Boltzmann transport theory serves as proof of the validity of our description of the hydrodynamic behavior of charge current at CNP.

{\it Kinetic theory approach.}

Electronic transport in the hydrodynamic regime is dominated by momentum-conserving electron-electron collisions\footnote{Umklapp processses are neglected} with characteristic time scale $\tau_{{ee}}$ \cite{Kovtun_2012,Lucas2018,Narozhny2019,Fritz2024}. We assume that: 1) diffusive mechanisms have significantly larger timescales; 2) quasi-particles remain well-defined even in the presence of interactions. Transport properties of the system are thus described by the Boltzmann equation :
\begin{equation}
\left (\frac{\partial }{\partial t} + \vec{v}_\lambda\cdot \nabla_{\vec{r}}  -e\vec{E}\cdot \nabla_{\vec{k}}  \right)f_\lambda(\vec{k}, \vec{r}, t)=  \mathcal{I}^{\text{ee}}_{\lambda}.
\label{eq:BoltzmannEq}
\end{equation}
Here $f_{\lambda}(\vec{k}, \vec{r}, t)$ defines the distribution function for the electrons. It depends on space $\vec{r}$, momentum $\vec{k}$, and time $t$ with $\lambda = \pm$ being the band index ($+$ ($-$) refers to the conduction (valence) band). Corresponding energies and velocities are: $\epsilon_{\pm} = \pm v_F k$ and $\vec{v}_{\pm} = \vec{\nabla}_{\vec{k}} \epsilon_{\pm} = \pm v_F \vec{k}/k$ (we consider only Dirac dispersion relation). In thermal equilibrium $f_{\pm} = (e^{\pm v_F k/T}+1)^{-1}$ at CNP.   In order to derive the current flow profile, we adopt a relaxation-time approximation for the collision integral \cite{Pongsangangan2022}: 
\begin{equation}
\label{eq:relaxatimeapprox}
\mathcal{I}^{\text{ee}}_\lambda = - \frac{\delta f_\lambda - \delta f_{-\lambda}}{\tau_{ee}}.
\end{equation}
The mean free time $\tau_{ee}$ accounts for the scattering between electrons of different bands and $\delta f_\lambda$ is the perturbation of the distribution functions, satisfying $\delta f_+ = -\delta f_-$ at CNP.  

\begin{figure}[]
  \centering
\includegraphics[width=0.5\textwidth , angle=0]{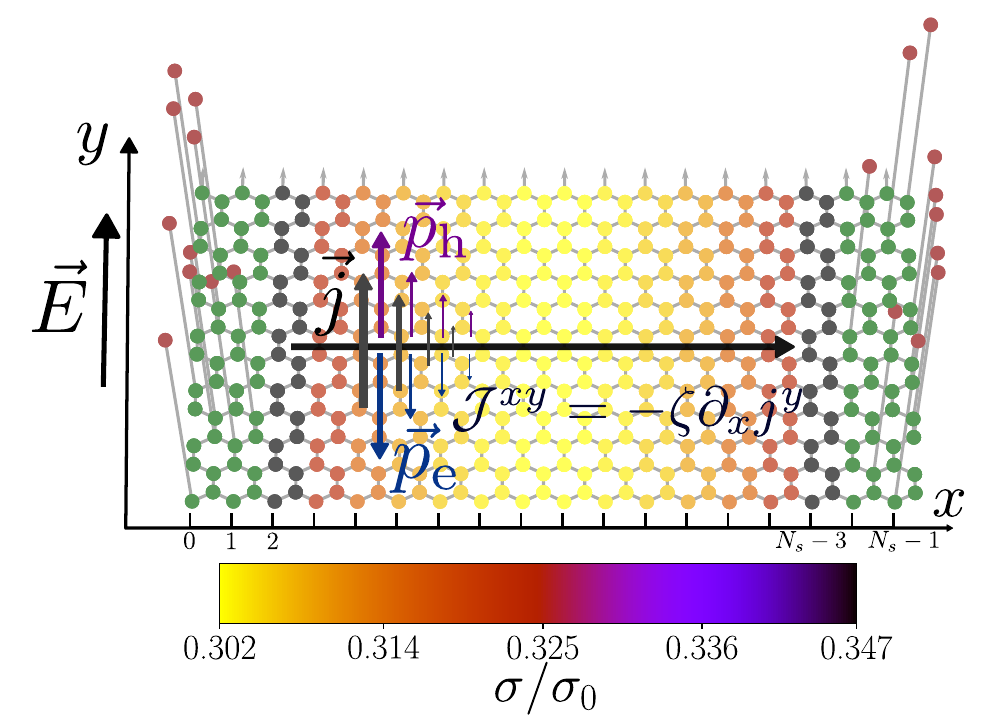}
        \caption{Scheme of the smallest simulated sample consisting of 18 stripes of carbon atoms. We impose periodical boundary conditions in y-direction, and open boundaries with disorder (adatoms, shown in red) in x-direction. We place 8 (2) adatoms randomly on the edges (shown in green). The electric field $\vec{E} = E \vec{e}_y$ drives electron and hole fluids in the opposite directions. While the momentum density remains zero, the current distribution has a non-trivial gradient $\vec{\nabla}\vec{j} \ne 0$. As a result, there exists a current flux $\mathcal{J}^{xy}$ diffusing from the region with high current density to the region with small current density. This results in the conductivity profile, visible through the heatmap corresponding to Fig~\ref{fig:2ProfilesSACWithoutInset}\textcolor{red}. The green sites on the edges correspond to the stripes with disorder and are not taken into account when fitting the conductivity profiles. $\sigma_0=1/4$  in the units of $e^2/\hbar$. }
  \label{fig:lattice}
\end{figure}

The equation for charge current is obtained from the Boltzmann equation \eqref{eq:BoltzmannEq} by multiplying both sides by $N e\vec{v}_\pm$ with $N=4$ accounting for valley and spin degeneracy and integrating over momentum:  
\begin{equation}
\label{eq:currentequation}
\partial_t {j}^n + \partial_m {\mathcal{J}}^{nm} +e {E}^m  {\mathcal{E}}^{mn} = -\frac{{j}^n}{\tau_{ee}/2}.
\end{equation}
Note that there is a factor 2 different in the relaxation time between Eqs.\eqref{eq:relaxatimeapprox} and \eqref{eq:currentequation}. This stems from the particle-hole symmetry that gives $\vec{v}_\lambda = -\vec{v}_{-\lambda}$. Thus, 
\begin{align}
& Ne\sum_{\lambda}\int d{\vec{k}} \vec{v}_\lambda \mathcal{I}^{\text{ee}}_\lambda \\ &= - \frac{1}{\tau_{ee}}Ne\sum_{\lambda} \int d{\vec{k}}\left(\vec{v}_{\lambda}\delta f_\lambda +  \vec{v}_{-\lambda}\delta f_{-\lambda}\right)\nonumber = -\frac{2 \vec{j}}{\tau_{ee}},
\end{align}
(see SM \cite{Supplement} for the detailed derivation).
Here $m,n \in \{x,y\}$ denote the direction of the vectors. 
$\vec {j} = \vec {j}_+ + \vec {j}_-$ is the total charge current density and the current densities of electrons ($\vec{j}_+$) and holes ($\vec{j}_-$) are defined as $\vec{j}_{\pm} = N e\int d{\vec{k}} \vec{v}_{\pm}  \tilde{f}_{\pm} $. Because we keep the charge $e$ positive within our formalism, the redefined distributions are $\tilde{f}_{+} = f_+$  and $\tilde{f}_- = f_--1$. This formalism is equivalent to keeping the distribution functions positive and changing the sign of the holes charge. Since $\tilde f_+ + \tilde f_- =0$ at half filling, and velocities for both bands have different signs, currents $\vec {j}_\lambda$ for electrons and holes add up (see also Fig. \ref{fig:lattice}).  This redefinition amounts to a change from the two-band basis to the electron-hole basis. As a result of this, we have a two-fluid model consisting of electrons and holes as elementary degrees-of-freedom of the system. The current flux tensor ${\mathcal{J}}^{nm}= {\mathcal{J}}^{nm}_+ +{\mathcal{J}}_-^{nm}$ plays a role comparable to the stress-energy tensor in the NS equation. The tensor $\mathcal{E}^{mn} = \mathcal{E}^{mn}_+ + \mathcal{E}^{mn}_- $ is coupled to the external electric field. These tensors are defined as $\mathcal{J}^{nm}_{\pm} = N e\int d{\vec{k}} {v}^{n}_{\pm} {v}^m_{\pm} \tilde{f}_{\pm}$ and
$\mathcal{E}^{mn}_{\pm} = N e\int d{\vec{k}} \partial_{m}{v}^{n}_{\pm} \tilde{f}_{\pm}$. It can be shown that ${\mathcal{E}^{mn}}$ is diagonal for Dirac fermions: ${\mathcal{E}^{mn}} =  -(2Te\ln 2)/\pi \delta^{mn} \equiv -e\mathcal{E} \delta^{mn}$, remaining non-zero at CNP and providing a leverage for external electric field to induce a charge current in the system. 

As a sanity check, we solve the current equation \eqref{eq:currentequation} for the homogeneous and stationary solution. In agreement with previous results \cite{Pongsangangan2022, Fritz_2008}, it gives the critical conductivity $\sigma_Q = e\mathcal{E}\tau_{ee}/2=  e^2(T\ln 2)\tau_{ee}/\pi $. 

In this paper,we focus on the inhomogeneous current profile when a current gradient is present, i.e. $\nabla_xj^y(x) \neq 0$.  As a result of the current gradient, a current flux $\mathcal{J} ^ {xy}$ flows from the region with high current density to low current density as shown in Fig.\ref{fig:lattice}. 
Phenomenologically, we assume a linear relationship  between the current flux tensor and the gradient of the current: 
\begin{equation}
\label{eq:constitutiverelation}
\mathcal{J} ^ { xy }  = -{\zeta} \nabla_x {j}^{y}.
\end{equation}
The current diffusion coefficient $\zeta$ represents a material's ability to conduct a current flux flowing in $x-$direction to transfer and, as a result, equilibrate the $y$-current in between layers of the fluid (see Fig. \ref{fig:lattice}) (compare it with kinematic viscosity which represents the conductivity of momentum flux 
in the presence of a momentum density gradient \cite{Kovtun_2012}).
We emphasize that $\zeta$ is distinct from the kinematic viscosity at CNP due to complete decoupling of the momentum density from the electric current. Later, the linear connection (\ref{eq:conductivityprofileforfit}) is confirmed when both $\mathcal{J} ^ { xy }$ and $\nabla_x {j}^{y}$ are computed using the distribution functions resulting from the solved linearized Boltzmann equation. It allows us to estimate $\zeta$ in the kinetic theory.

The final stationary ($\partial_t \vec{j}=0$) current equation is
\begin{equation}
-{\zeta} \nabla_x^2 {j}^y(x) +e{E}^y {\mathcal{E}}^{yy} = -\frac{{j}^y}{\tau_{ee}/2},
\label{eq:NStype}
\end{equation}
augmented by the continuity equation $\vec{\nabla} \cdot \vec{j} = 0$. Note the similarity with the NS equation for velocity. 
If we solve it for a strip of width $w$, oriented along the y-axis (same direction as the electric field), we get the catenary curve
\begin{equation}
j^y(x) = \left(\sigma_Q + \sigma_B \frac{\cosh \left( A x \right)}{\cosh \left( A \frac{w}{2}\right)}   \right) E_y,
\label{eq:conductivityprofileforfit}
\end{equation}
where $A=\sqrt{(e\mathcal{E})/(\zeta \sigma_Q)}$, and $\sigma_B$ is defined by the boundary conditions.

We now return to the full Boltzmann equation to estimate $\zeta$ :
\begin{equation}
\vec{v}_\lambda\cdot \nabla_{\vec{r}}  f_\lambda(\vec{k}, \vec{r}, t)=  \mathcal{I}^{\text{col}}_{\text{ee}}.
\label{eq:Boltzmann1}
\end{equation} 
Here we turn to the more exact collision integral that is obtained from Fermi's golden rule (see SM \cite{Supplement} for the full expression). The way in which this collision integral reduces to that from the relaxation-time approximation in Eq.\eqref{eq:relaxatimeapprox} is discussed in SM \cite{Supplement}.
 The above Boltzmann equation was previously solved up to the leading-log approximation by the variational method \cite{Ziman2001} for  shear viscosity \cite{Fritz2009,Kitinan2024} of graphene. Applying the same approach, we arrive at the following expression for $\zeta$ in the leading-log approximation:
\begin{equation}
\frac{\zeta}{a^2t} = C \left( \epsilon a t\right)^2 \frac{t}{T},
\label{eq:zeta_T}
\end{equation}
with dimensionless $C=25.76$.  $t$ is the C-C hopping, $\epsilon$ is dielectric permittivity of surrounding media and $a$ is the unit lattice vector length in graphene. Eq. \eqref{eq:zeta_T} is strictly speaking valid at low temperatures where the dispersion relation is approximately linear.

{\it Microscopic description and QMC simulations}

In order to demonstrate hydrodynamics of charge current at the CNP in numerically exact QMC simulations, we consider free standing graphene strips similar to the one shown in Fig.~\ref{fig:lattice}, with $N_s$ stripes in x-direction. Some sort of boundary conditions are needed to directly observe non uniform hydrodynamic flow profiles. Since the lattice size is limited in QMC simulations, we can not consider broken edges like in \cite{Kiselev2019}. The sign problem would appear if we try to impose boundary conditions by gating the edges of the sample. Thus, the only choice to introduce disorder is to randomly place adatoms on top of the sites along the two edges, without violating the particle-hole symmetry (see Fig.~\ref{fig:lattice}). While both approaches (gating and broken edges) are certainly more relevant experimentally, it only matters for the observation of the hydrodynamic behavior, whether the non-uniform flow profiles follow the NS-type equation (\ref{eq:NStype}). The origin of the spatial non-uniformity of the current is of secondary importance.  We refer to the regions where disorder is introduced by random placement of adatoms as edges. These regions are highlighted in green in Fig.~\ref{fig:lattice} and correspond to the stripes at $ x = 0,1, N_s-1$ and $N_s-2$. The hydrodynamic boundary layers are defined as the first stripes adjacent to the disordered regions. They are located at $ x = 2, N_s-3$ and represent the first disorder-free stripes.

We simulate the following particle-hole symmetric quantum Hamiltonian with long-range Coulomb interactions (LRC)
\begin{align}
    \hat H_{\mathrm{LRC}} &=-\sum_{<i,j>,\sigma} t ( \hat a^\dag_{i, \sigma} \hat a_{j, \sigma} + \mathrm{h.c.} ) 
     + \frac{1}{2} \sum_{ij} V_{ij} \hat n_{i} \hat n_{j},
      \label{eq:Qham}
\end{align}
where $<i,j>$ are nearest neighbor sites. $\hat  a_{i, \sigma}$ are annihilation operators for electrons with spin $\sigma=\uparrow, \downarrow$ on site $i$ and $V_{ij}$ is a matrix of unscreened Coulomb interactions between the charge carriers described by the charge density operator $\hat n_{i} =     \sum_{\sigma} \left(  \hat{a}^{\dag}_{i,\sigma} \hat{a}^{}_{i,\sigma}  -  \frac{1}{2} \right) $.  The hopping is $t = 2.7 \eV$ and the matrix of potentials for free-standing graphene is taken from \cite{Wehling_2011} (see SM \cite{Supplement} for more details). The on-site interaction equals to 9.3 eV, nearest-neighbor interaction is 5.5 eV, and the long-rang Coulomb tail corresponds to the vacuum dielectric permittivity. These parameters were previously validated by comparing QMC calculations with experimental data for the renormalized Fermi velocity $v_F$ \cite{Ulybyshev:2021xao}.

  The hopping between adatoms $\hat c^\dag_{i, \sigma}$ and regular sites $ \hat a^\dag_{i, \sigma}$, is described by a particle-hole symmetric hybridization Hamiltonian $\hat H_{\mathrm{adatom}} = \sum_{i,\sigma} t' ( \hat a^\dag_{i, \sigma} \hat c_{i, \sigma} + \mathrm{h.c.} ) $, with $t' = 10t$, roughly corresponding to hydrogen adatoms on top of carbons \cite{PhysRevLett.114.246801}. The full microscopic Hamiltonian of the system is therefore $\hat H = \hat H_{\mathrm{adatom}} + \hat H_{\mathrm{LRC}} $.

The Hamiltonian is simulated using the finite temperature auxiliary field Quantum Monte Carlo method \cite{Blankenbecler81,White89,Assaad08_rev,Ulybyshev13,Hohenadler14}, specifically its realization from the Algorithm for Lattice Fermions (ALF) \cite{ALF_v2}. In order to compute the current profile, we measure the diagonal element of the stripe $x$-resolved DC conductivity tensor $\sigma_{yy}(x,\omega=0)$ (averaged over the $y$-coordinate). It is obtained using the stochastic
analytical continuation method (SAC) \cite{Beach04a,Sandvik98,SHAO20231} via Green-Kubo relations from the Euclidean current-current correlator \cite{Mahan90,Kubo57,Jarrell96,PhysRevLett.118.266801}. In order to justify the stability of the SAC method, we reproduced the same results from the middle-point of the current-current correlator \cite{Trivedi1996} and performed additional tests using exact free conductivity (see SM \cite{Supplement} for further details).

 \begin{figure}[]
  \centering
\includegraphics[width=0.45\textwidth , angle=0]{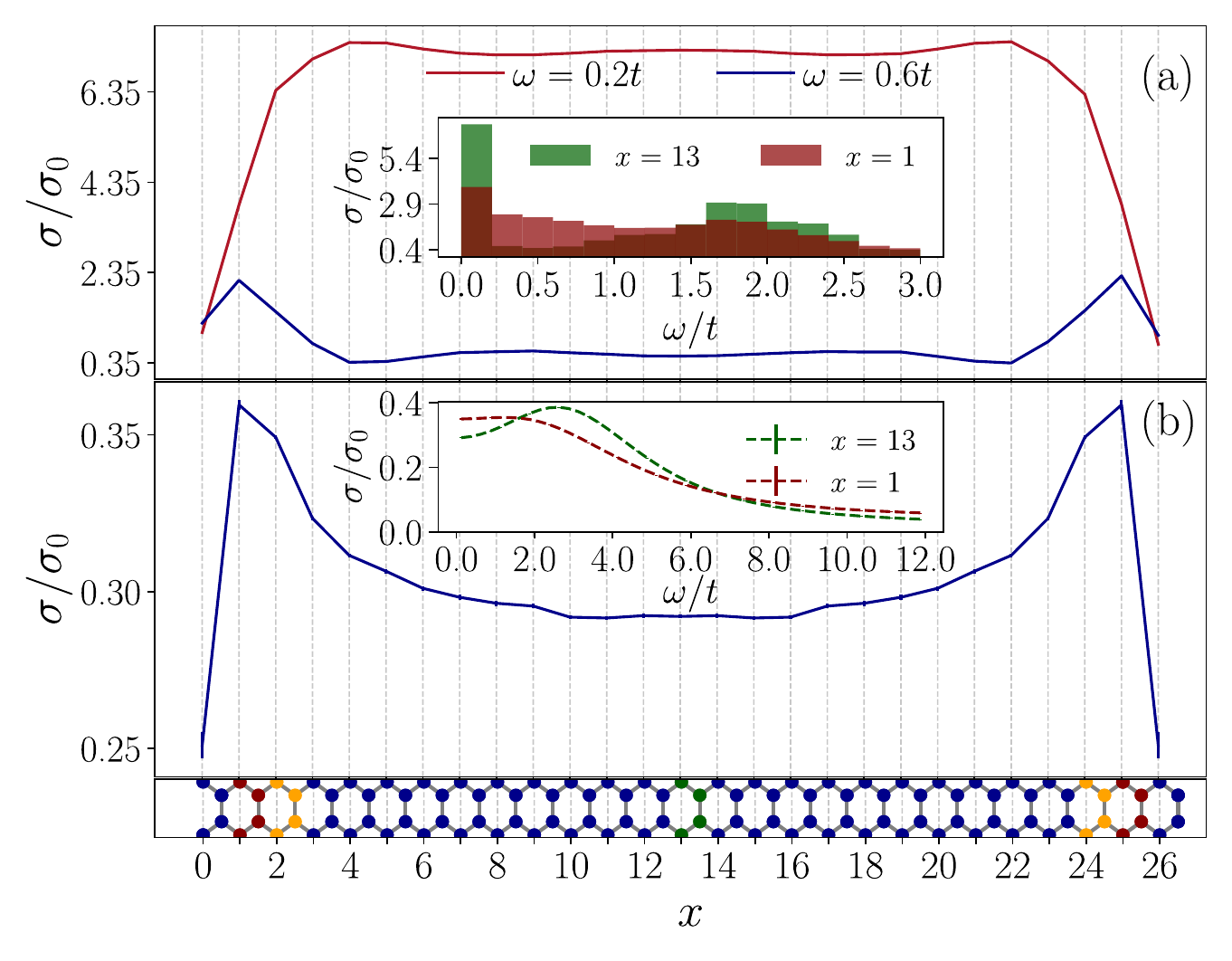}
        \caption{(a) Local conductivity profiles at two different frequencies for the free tight-binding model with randomly distributed adatoms at the edges at a temperature $T=0.67t$. The inset shows $\sigma(\omega)$ profiles for the same free Hamiltonian in the middle of the sample ("bulk") and at the edge. Since the  conductivity is a series of delta-functions in the free case, each bar in the plot shows the sum of weights of these delta functions in the corresponding interval of frequencies. (b) Local DC conductivity profile across the sample in the interacting case for the same lattice and temperature. SAC data, obtained from the Euclidean current-current correlator and with the inset showing the boundary and bulk $\sigma(\omega)$. Below the figure, we sketched the cross-section of the lattice with edge stripe, bulk stripe and hydrodynamic boundary highlighted in different colors. }
  \label{fig:FreeVSInteractingProfilesOmega}
\end{figure}

 \begin{figure}[]
  \centering
\includegraphics[width=0.45\textwidth , angle=0]{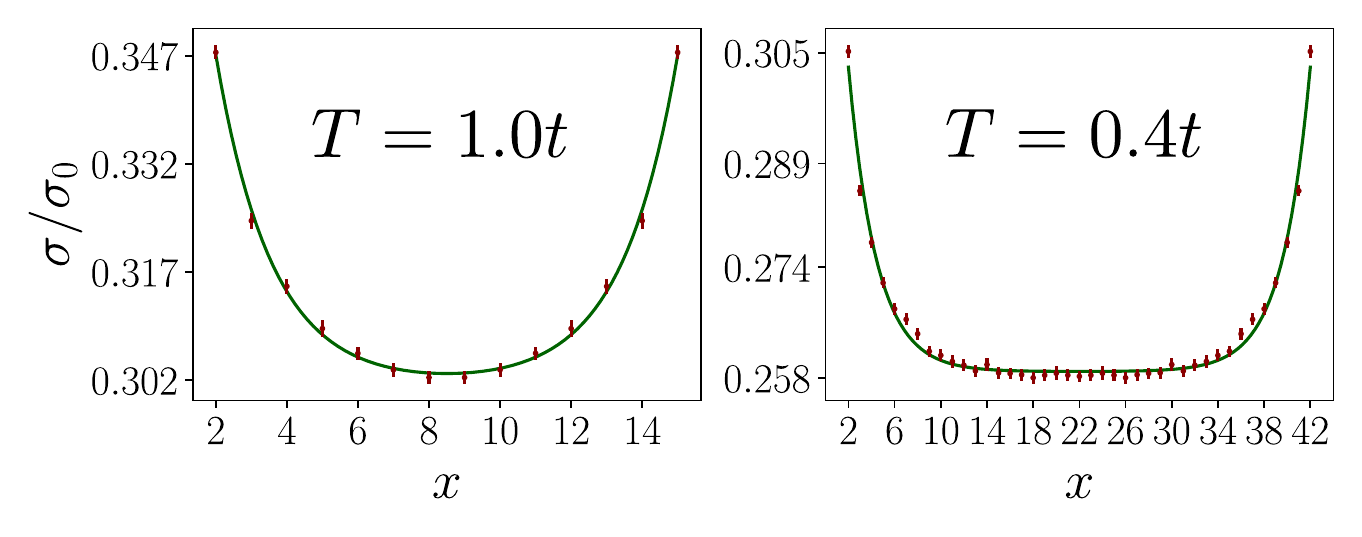}
        \caption{The stripe resolved DC conductivity profiles across the sample in the interacting case for two different temperatures. SAC data, obtained from the Euclidean current-current correlator. More profiles are in SM \cite{Supplement}.}
  \label{fig:2ProfilesSACWithoutInset}
\end{figure}

We start from the analysis of the boundary conditions imposed by the adatoms at the edges. Frequency dependent profiles of the free conductivity obtained from the tight-binding part of the full Hamiltonian $\hat H$ show that the imposed boundary conditions are substantially different for different frequencies (Fig.~\ref{fig:FreeVSInteractingProfilesOmega}\textcolor{red}{a}).
Small frequencies (Drude peak regime) have almost no-slip type of boundary conditions,  with the conductivity sharply decreasing at the edges, while optical frequencies around the Dirac plateau ($\omega\approx 0.5t$) have reverse boundary conditions with a conductivity at the edges larger than in the bulk. In the interacting and thus hydrodynamic case (Fig.~\ref{fig:FreeVSInteractingProfilesOmega}\textcolor{red}{b}), the Drude peak disappears from the $\sigma(\omega)$ profiles due to a large electron-electron scattering rate, and we are left with the Dirac plateau regime extending to small frequencies, a situation already observed in \cite{PhysRevB.94.085421, PhysRevLett.118.266801}. Thus the boundary conditions for small frequencies in the strongly interacting regime are similar to the larger frequency regime of the free case with the current falling starting from the hydrodynamic boundary layer towards the bulk. Hence we fit our QMC profiles in the bulk using the catenary curve \eqref{eq:conductivityprofileforfit} with $\sigma_B>0$.

 \begin{figure}[]
  \centering
\includegraphics[width=0.45\textwidth , angle=0]{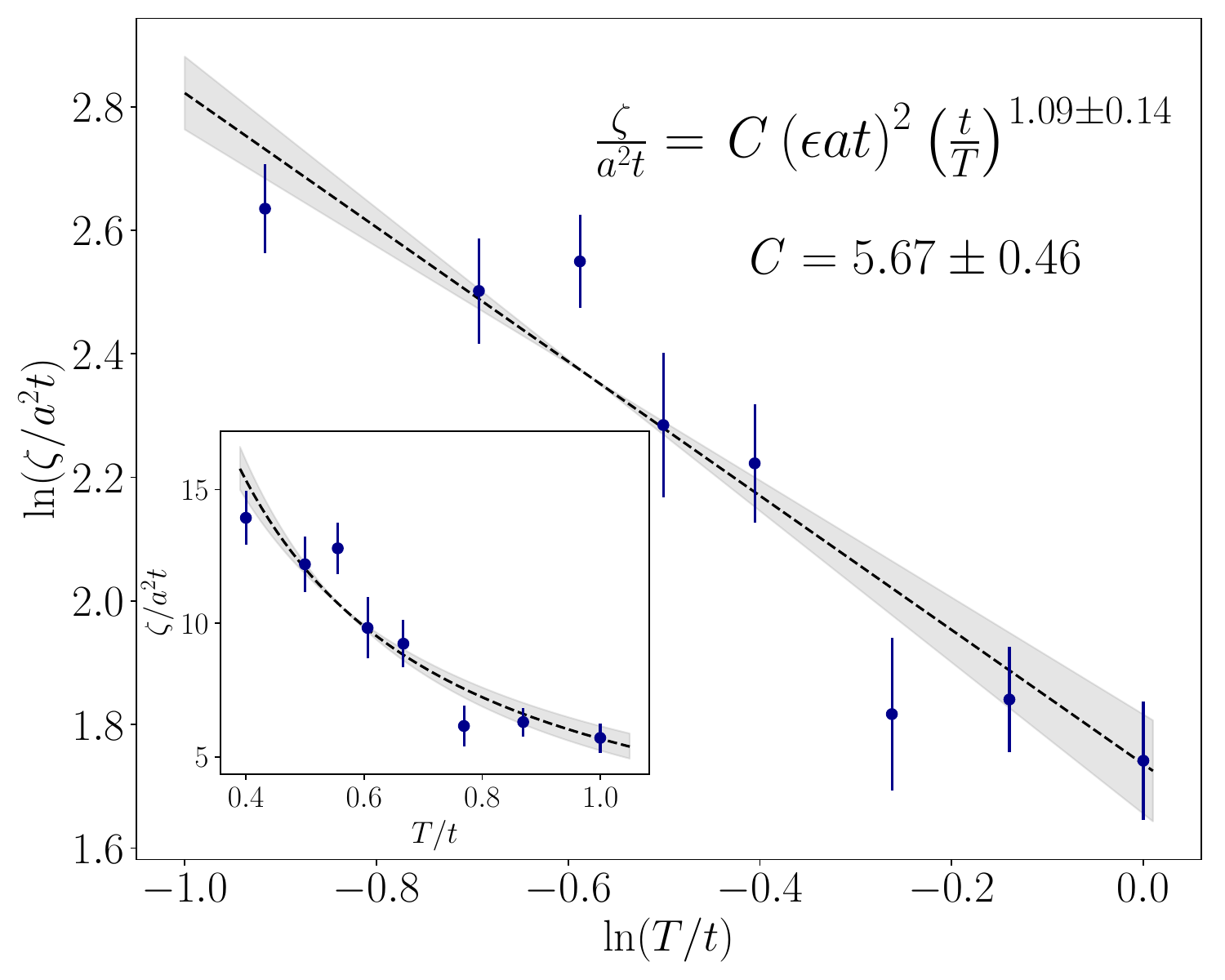}
        \caption{Temperature dependent current diffusion coefficient $\zeta$ in log-log representation (main plot) and with linear scales (inset). Fit is shown with the dotted line and the error margin of the fitting function is shown in grey. $\epsilon = 1$ for QMC simulations of free-standing graphene.}
  \label{fig:CurrentDiffusionCoefficientFit}
\end{figure}

The conductivity profiles computed for the full interacting Hamiltonian, after averaging over various adatom configurations, are displayed in Fig.~\ref{fig:2ProfilesSACWithoutInset}\textcolor{red}. In order to have a high density of temperature excitations for our small lattice sizes (in another words, to satisfy the condition $\omega \ll T$ for hydrodynamic regime \cite{Fritz_2008}), the simulations were carried out for high temperatures, and the width of the sample is increased inversely proportional to the lowering $T$. The length in y-direction is kept constant. It only influences the resolution for the y-component of momentum. However, this resolution should be compared with the temperature scale. Since the temperature is high, we have enough levels within the $(0, T)$ interval of energies, and an increase of resolution for the y-component of momentum will not substantially alter the results. The data reproduces remarkably well the expected conductivity profiles of Eq.~\ref{eq:conductivityprofileforfit}. From them, we extract the values of $\zeta$,  also computing the diagonal elements of  the tensor ${\mathcal{E}}$ using the exact interacting Green Functions obtained in QMC (see SM \cite{Supplement} for the details). The resulting temperature dependent $\zeta$ is presented in Fig.~\ref{fig:CurrentDiffusionCoefficientFit}. Fitting at $\epsilon=1$ yields the same power law function as in  \eqref{eq:zeta_T}: $\zeta \sim 1/T^{{ {1.09} \pm {0.14} }}$, but  
with $C = 5.67 \pm 0.46$. The difference between the theoretical expectation and QMC results can be attributed to the approximations employed in kinetic theory: 1) Dirac dispersion relation instead of the full expression for the honeycomb lattice; 2) Only tree-level diagrams in the collision integral in \eqref{eq:Boltzmann1}. The latter deficiency can be also connected to the importance of higher-order diagrams in the  comparison of $v_F$ renormalization with the perturbative RPA result reported in \cite{Ulybyshev:2021xao}. It seems that higher-order corrections to the scattering amplitudes play an important role in the transport properties of free-standing graphene.

{\it Conclusion.}

We demonstrated the emergence of hydrodynamic charge transport in freestanding graphene at CNP in the absence of momentum flow.  First, we obtained charge current profiles using unbiased QMC simulations in a controlled environment, where momentum-conserving particle-particle collisions are the sole scattering mechanism in the bulk, while diffusive scattering is confined to the edges to establish boundary conditions. Second, we developed a  description for the charge current based on a two-fluid (electrons and holes) Boltzmann formalism.

This description leads to NS-type equations written directly for the charge current with the current diffusion coefficient $\zeta$ replacing viscosity. The current profiles obtained from QMC align well with the solutions of these equations, allowing us to extract the temperature dependence of  $\zeta(T)$ from these fits. These results show qualitative agreement with Boltzmann transport theory. We expect a similar picture for any system with particle-hole symmetry at half-filling, e.g. for AB-stacked bilayer graphene with quadratic band touching or for any narrow-gap semiconductor with symmetric bands.

Another interesting question is the numerical accuracy of the transport coefficients obtained from the Boltzmann equation. Our paper demonstrates, how QMC can serve as a replacement of experiments in sourcing transport data to check and improve approximations used in the kinetic theory. The main advantage of this technique is the precise control over scattering mechanisms in microscopic simulations. Using this approach, we are planning to further investigate possible deficiencies in the computation of the collision integral for strongly correlated quantum systems.  
\\
\noindent\textit{The data that support the findings of this article are openly available in \cite{Data}}

\begin{acknowledgments}

KP thanks former collaborations and discussions
with L. Fritz, H.T.C. Stoof, T. Ludwig, S. Grubinskas, P. Cosme, E. Di Salvo, T. Meng.
MU thanks J. Schmalian for discussions about the boundary conditions and AR thanks R. Meyer for insightful exchanges about hydrodynamics in QFT.\\ \indent
MU (AR)  thanks  the  DFG   for financial support  under the projects UL444/2-1, Project number 495044360 (AS120/19-1, Project number 530989922). 
FFA acknowledges financial support from the DFG through the W\"urzburg-Dresden Cluster of Excellence on Complexity and Topology in Quantum 
Matter - \textit{ct.qmat} (EXC 2147, Project No.\ 390858490)   as  well as  the SFB 1170 on Topological and Correlated Electronics at Surfaces and Interfaces (Project No.\  258499086). KP acknowledges funding by the Deutsche Forschungsgemeinschaft (DFG) via the Emmy Noether Programme (Quantum Design grant, ME4844/1, project- id 327807255), project A04 of the Collaborative Research Center SFB 1143 (project-id 247310070), and the Cluster of Excellence on Complexity and Topology in Quantum Matter ct.qmat (EXC 2147, project-id 390858490).
We gratefully acknowledge the Gauss Centre for Supercomputing e.V.\ (www.gauss-centre.eu) for funding this project by providing computing time for the computation of the current-current correlator on the GCS Supercomputer SUPERMUC-NG at the Leibniz Supercomputing Centre (www.lrz.de, project number pn73xu), as well as the scientific support and HPC resources provided by the Erlangen National High Performance Computing Center (NHR@FAU) of the Friedrich-Alexander-Universit\"at Erlangen-N\"urnberg (FAU) under the NHR project b133ae to carry out the SAC analysis. NHR funding is provided by federal and Bavarian state authorities. NHR@FAU hardware is partially funded by the German Research Foundation (DFG) -- 440719683.  JUWELS ~\cite{JUWELS} supercomputer was used for the calculation of $\mathcal{E}$ tensor. The numerical calculations were carried out with the Algorithms for Lattice Fermions (ALF) library \cite{ALF_v2}. 
\nocite{PhysRevLett.103.025301}
\nocite{Buividovich12}
\nocite{PhysRevB.110.155120}

\end{acknowledgments}

\begin{widetext}
\section*{ Supplementary Material}

\section{Particle hole symmetry and the decoupling of charge and momentum currents.}

Our aim is to show that in the linear response regime the electric field does not  generate any momentum response  but does generate a current response.  To define particle-hole symmetric Hamiltonians, it is 
convenient to  consider a Bravais lattice with two orbitals per unit cell. We will work with the spinor 
\begin{equation}
    \hat{\Psi}^{\dagger}_{\ve{i},\sigma} =  \left( \hat{a}^{\dagger}_{\ve{i},\sigma}, \hat{b}^{\dagger}_{\ve{i},\sigma} \right) 
\end{equation}
where $\ve{i}$ denotes the unit cell, $\sigma$ the spin and $a,b$  the orbital index.  The particle-hole transformation that we will consider reads: 
\begin{equation}
\hat{T}^{-1} \hat{\Psi}^{\dagger}_{\ve{i},\sigma}  \hat{T} = \tau_z \hat{\Psi}^{T}_{\ve{i},\sigma} 
\equiv 
\left( 
  \hat{a}^{\phantom\dagger}_{\ve{i},\sigma},  
  -\hat{b}^{\phantom\dagger}_{\ve{i},\sigma} \right)
\end{equation}
where $\tau_z= \text{diag}(1,-1)$ is a Pauli matrix acting on the orbital index. 
The model Hamiltonian $\hat{H}$ is particle-hole symmetric if, 
\begin{equation}
  \left[ \hat{H},  \hat{T} \right] = 0. 
\end{equation}
This is certainly the case for the  graphene Hamiltonian considered here.
To define the current operator we include a vector potential on bonds $\ve{b}$,  $\ve{A}_{\ve{b}}$,  via the Peierls substitution and the current operator is defined by: 
\begin{equation}
    \hat{J}^{\alpha}_{\ve{b}} (\ve{A}) = \frac{\partial \hat{H}(A)}{\partial A^{\alpha}_{\ve{b}}}
\end{equation} 
In the presence of the vector potential field, 
\begin{equation}
       \hat{T}^{-1} \hat{H}(A)  \hat{T}= \hat{H}(-A)
\end{equation}
such that 
\begin{equation}
  \hat{T}^{-1} \hat{J}^{\alpha}_{\ve{b}}(\ve{A})   \hat{T}=  -\hat{J}^{\alpha}_{\ve{b}}(-\ve{A}).
\end{equation}
In particular, the paramagnetic current,  $\hat{J}^{p,\alpha}_{\ve{b}} \equiv \hat{J}^{\alpha}_{\ve{b}}(\ve{0}) $, satisfies 
\begin{equation}
  \hat{T}^{-1} \hat{J}^{p,\alpha}_{\ve{b}}   \hat{T}=  -\hat{J}^{p,\alpha}_{\ve{b}}
\end{equation}
and is hence odd under particle-hole symmetry.

Momentum  $\hat{\ve{P}} $ is the generator  of translations:
\begin{equation}
 e^{i \hat{\ve{P}}\cdot \ve{r}} \Psi^{\dagger}_{\ve{i},\sigma} e^{-i \hat{\ve{P}}\cdot \ve{r}} = \Psi^{\dagger}_{\ve{i}+\ve{r},\sigma}.  
 \label{eq.defmomentum}  
\end{equation}
Acting on both sides of the  above equation with the particle-hole transformation gives: 
\begin{eqnarray}
  & & \phantom{\leftrightarrow} e^{i \hat{T}^{-1}\hat{\ve{P}} \hat{T} \cdot \ve{r}} \hat{T}^{-1}\Psi^{\dagger}_{\ve{i},\sigma} \hat{T} e^{-i \hat{T}^{-1}\hat{\ve{P}}\hat{T}\cdot \ve{r}} = \hat{T}^{-1}
  \Psi^{\dagger}_{\ve{i}+\ve{r},\sigma} \hat{T} \nonumber  \\
  & & \leftrightarrow  e^{i \hat{T}^{-1}\hat{\ve{P}} \hat{T} \cdot \ve{r}} 
  \tau_z\Psi^{T}_{\ve{i},\sigma}  e^{-i \hat{T}^{-1}\hat{\ve{P}}\hat{T}\cdot \ve{r}} = 
  \tau_z\Psi^{T}_{\ve{i}+\ve{r},\sigma} \nonumber \\
  & & \leftrightarrow  e^{i \hat{T}^{-1}\hat{\ve{P}} \hat{T} \cdot \ve{r}} 
  \Psi^{T}_{\ve{i},\sigma}  e^{-i \hat{T}^{-1}\hat{\ve{P}}\hat{T}\cdot \ve{r}} = 
  \Psi^{T}_{\ve{i}+\ve{r},\sigma}  
\end{eqnarray}
Since from Eq.~\ref{eq.defmomentum}   follows
$e^{i \hat{\ve{P}}\cdot \ve{r}} \Psi^{T}_{\ve{i},\sigma} e^{-i \hat{\ve{P}}\cdot \ve{r}} = \Psi^{T}_{\ve{i}+\ve{r},\sigma}$ we have that:
\begin{equation}
  e^{i \hat{T}^{-1}\hat{\ve{P}} \hat{T} \cdot \ve{r}} 
  \Psi^{T}_{\ve{i},\sigma}  e^{-i \hat{T}^{-1}\hat{\ve{P}}\hat{T}\cdot \ve{r}}   =e^{i \hat{\ve{P}}\cdot \ve{r}} \Psi^{T}_{\ve{i},\sigma} e^{-i \hat{\ve{P}}\cdot \ve{r}}  \, \,\, \,   \forall \, \, \, 
  \ve{r},
\end{equation}  
such that 
\begin{equation}
  \hat{T}^{-1}\hat{\ve{P}} \hat{T} = \hat{\ve{P}}.
\end{equation}
Since  the paramagnetic current (momentum) is odd (even)  under  particle-hole symmetry  we have that
\begin{equation}
    \text{Tr}  \left(  e^{-\beta \hat{H} }  \left[ \hat{P}^{\alpha}(t), \hat{J}^{p,\beta}_{\ve{b}} \right]  \right)= 0
\end{equation} 
Thereby, the electric field generates a current flow but not momentum transport within linear response theory for particle-hole symmetric Hamiltonians.

\section{Kinetic theory approach and the calculation of the current diffusion coefficient}

\subsection{Relaxation-time approximation}
The relaxation-time approximation used in the main text is obtained by linearizing the collision integral. To this end, we write $f_\lambda(\vec{k}) = f^0_\lambda(\vec{k}) + \delta f_\lambda(\vec{k})$ with $f^0_\lambda$ being the Fermi-Dirac function for electrons in equilibrium. This gives
 
 \begin{eqnarray}
\mathcal{I}^{\text{ee}}_{\lambda} &\approx& -\frac{2\pi}{v_F}\int_{\vec{k}_1,\vec{q}} f_\lambda^0(\vec{k})f^0_{-\lambda}(\vec{k}_1)(1-f^0_\lambda(\vec{k}-\vec{q}))(1-f^0_{-\lambda}(\vec{k}_1+\vec{q})) \nonumber \\ &&\delta\left(k-k_1-|\vec{k}-\vec{q}|+|\vec{k}_1+\vec{q}|\right) R_1(\vec{k},\vec{k}_1,\vec{q})\left( \delta f_\lambda(\vec{k}) + \delta f_{-\lambda}(\vec{k}_1) - \delta f_\lambda(\vec{k}-\vec{q}) - \delta f_{-\lambda}(\vec{k}_1+\vec{q})\right) \nonumber \\ && -\frac{2\pi}{v_F}\int_{\vec{k}_1,\vec{q}} f^0_\lambda(\vec{k}) f^0_\lambda(\vec{k}_1) (1-f^0_\lambda(\vec{k}-\vec{q}))(1-f^0_\lambda(\vec{k}_1+\vec{q})) \nonumber  \\ && \delta(k+k_1-|\vec{k}-\vec{q}|-|\vec{k}_1+\vec{q}|) R_2(\vec{k},\vec{k}_1,\vec{q}) \left(\delta f_{\lambda}(\vec{k}) + \delta f_\lambda(\vec{k}_1)  - \delta f_{\lambda}(\vec{k}-\vec{q}) - \delta f_{\lambda}(\vec{k}_1+\vec{q})\right). 
 \end{eqnarray}
 Furthermore, we will include only scattering processes of which the incoming particles have the same momentum that is $\vec{k} = \vec{k}_1$ and ignore other processes. Outgoing particles remain in equilibrium. This gives
\begin{eqnarray}
\mathcal{I}^{\text{ee}}_{\lambda} &\approx& -\frac{2\pi}{v_F A}\int_{\vec{q}} f_\lambda^0(\vec{k})f^0_{-\lambda}(\vec{k})(1-f^0_\lambda(\vec{k}-\vec{q}))(1-f^0_{-\lambda}(\vec{k}+\vec{q})) \delta\left(|\vec{k}-\vec{q}|+|\vec{k}+\vec{q}|\right) R_1(\vec{k},\vec{k},\vec{q})\left( \delta f_\lambda(\vec{k}) + \delta f_{-\lambda}(\vec{k})\right) \nonumber \\ && -\frac{2\pi}{v_F A}\int_{\vec{q}} f^0_\lambda(\vec{k}) f^0_\lambda(\vec{k}) (1-f^0_\lambda(\vec{k}-\vec{q}))(1-f^0_\lambda(\vec{k}+\vec{q})) \delta(k+k-|\vec{k}-\vec{q}|-|\vec{k}+\vec{q}|) R_2(\vec{k},\vec{k},\vec{q}) \left(\delta f_{\lambda}(\vec{k}) + \delta f_\lambda(\vec{k}) \right).\nonumber \\ 
\end{eqnarray}
At CNP, $\delta f_\lambda = -\delta f_{-\lambda}$. This simplifies the collision and we obtain
\begin{eqnarray}
\mathcal{I}^{\text{ee}}_{\lambda}  \approx -\frac{\delta f_\lambda(\vec{k})-\delta f_{-\lambda}(\vec{k})}{\tau_{ee}}
 \end{eqnarray}
 with
 \begin{equation}
 \frac{1}{\tau_{ee}(\vec{k})} = \frac{2\pi}{v_F A}\int_{\vec{q}} f^0_\lambda(\vec{k}) f^0_\lambda(\vec{k} (1-f^0_\lambda(\vec{k}-\vec{q}))(1-f^0_\lambda(\vec{k}+\vec{q})) \delta(k+k-|\vec{k}-\vec{q}|-|\vec{k}+\vec{q}|) R_2(\vec{k},\vec{k},\vec{q}). 
 \end{equation}
 Note that the relaxation time $\tau_{ee}$ obtained here depends on momentum $\vec{k}$. However, in the main text, we simplify it by assuming that the relaxation time is constant.

\subsection{The Equation for Charge Current}

\noindent This section is devoted to a more detailed derivation for the equation of charge current mentioned in the main text. The derivation starts from the Boltzmann equation:

\begin{align}
\label{eq:Boltzmannequation}
\partial_t f_{\pm} + \vec{v}_{\pm} \cdot \vec{\nabla} f_{\pm}- e \vec{E} \cdot \vec{\nabla}_{\vec{k}} f_{\pm} = - \frac{\delta f_+ - \delta f_-}{\tau_{ee}}.
\end{align}
Here $f_{\pm}$ defines the distribution function for electron in the conduction ($+$) and valence $(-)$ bands with the corresponding energy $\epsilon_{\pm} = \pm v_F k$ and velocity $\vec{v}_{\pm} = \vec{\nabla}_{\vec{k}} \epsilon_{\pm} = \pm v_F \vec{k}/k$. Let us emphasize that $\vec{v}_- = -\vec{v}_+$. In thermal equilibrium $f_{\pm} = (e^{\pm v_F k/T}+1)^{-1}$ at charge neutrality. The mean free time $\tau_{ee}$ appears due to scattering processes between electrons and holes, and the collision integral is treated in treated in the relaxation time approximation. 

The equation of charge current is obtained from the Boltzmann equation \eqref{eq:Boltzmannequation} by multiplying it by $e\vec{v}_\pm$ followed by integrating over momenta. This gives
\begin{equation}
\label{eq:suppelectronholecurrents}
\partial_{t} \vec{j}_{\pm} + \vec{\nabla} \cdot \vec{\vec{\mathcal{J}}}_{\pm} + e \vec{E} \cdot \vec{\vec{\mathcal{E}}}_{\pm} = -\frac{ \vec{j}_+ + \vec{j}_-}{\tau_{ee}}.
\end{equation}
The current densities for electrons ($\vec{j}_+$) and holes ($\vec{j}_-$) are defined as $\vec{j}_{\pm} = e\int_{\vec{k}} \vec{v}_{\pm}  \tilde{f}_{\pm} $. While the distribution function for conduction-band electrons remains unchanged: $\tilde{f}_{+} = f_+$, the distribution function for electrons in the valence band is redefined by subtracting the infinite contribution of the Dirac sea $\tilde{f}_- = f_--1$. In addition, we defined a shorthand notation for second rank tensors
$\vec{\vec{\mathcal{J}}}_{\pm} = e\int_{\vec{k}} \vec{v}_{\pm} \vec{v}_{\pm} \tilde{f}_{\pm}$ and
$\vec{\vec{\mathcal{E}}}_{\pm} = e\int_{\vec{k}} \vec{\nabla}_{\vec{k}}\vec{v}_{\pm} \tilde{f}_{\pm}$. 

The electron and hole currents add up to the total current density: $\vec{j} = \vec{j}_+ + \vec{j}_-$.  The sum of Eq. \eqref{eq:suppelectronholecurrents} for both $\lambda$ values gives the equation of charge current:
\begin{equation}
\partial_t \vec{j} + \vec{\nabla}\cdot \vec{\vec{\mathcal{J}}} +e \vec{E} \cdot \vec{\vec{\mathcal{E}}} = -\frac{\vec{j}}{\tau_{ee}/2}.
\end{equation}

It should be noted that the second rank tensor $\vec{\vec{\mathcal{J}}}= \vec{\vec{\mathcal{J}}}_+ +\vec{\vec{\mathcal{J}}}_-$ is the counterpart of the stress-energy tensor in the Navier-Stokes equation and the second tensor $\vec{\vec{\mathcal{E}}} = \vec{\vec{\mathcal{E}}}_+ + \vec{\vec{\mathcal{E}}}_- $ is coupled to the external electric field. In contrast to momentum density governed by Navier-Stokes equation, at charge neutrality, current density is coupled to the electric field as $\vec{\vec{\mathcal{E}}}$ remains non-zero, i.e. for Dirac fermions, 
\begin{equation}
\vec{\vec{\mathcal{E}}} =  -(2Te\ln 2)/\pi \vec{\vec{1}} .
\end{equation}

 \subsection{Kinetic equation}

 In this section, we sketch the calculation of the current diffusion coefficient $\zeta$ from the kinetic theory approach reported in the main text.

A non-zero current density gradient, $\vec{\nabla}\vec{j}\ne0$  drives a system out of equilibrium. A stationary, but inhomogeneous, state of the system  obeys a Boltzmann equation reading as
\begin{equation}
\vec{v}_\lambda\cdot \nabla_{\vec{r}}  f_\lambda(\vec{k}, \vec{r}, t)=  \mathcal{I}^{\text{ee}}_\lambda
.
\end{equation}
Here $f_{\pm}$ defines the distribution function for Dirac electrons in the conduction ($+$) and valence $(-)$ bands with the corresponding energy $\epsilon_{\pm} = \pm v_F k$ and velocity $\vec{v}_{\pm} = \vec{\nabla}_{\vec{k}} \epsilon_{\pm} = \pm v_F \vec{k}/k$ close to the K point in the Brillouin zone. The collision integral $\mathcal{I}^{\text{col}}_{\text{ee}}$ was derived from a Schwinger-Keldysh quantum field theory and is written in terms of the Coulomb interaction up to $\alpha^2$ \cite{Fritz_2008}. Here $\alpha = e^2/6 \pi \hbar \epsilon a t$ defines effective fine structure constant of graphene, where $t$ is C-C hopping, $a$ the graphene unit cell length and $\epsilon$ is the dielectric constant defining the screening of the Coulomb potential by a substrate.   The corresponding collision integral describes $2\rightarrow2$ scattering and reads as
\begin{eqnarray}
\mathcal{I}^{\text{ee}}_\lambda &=& -\frac{2\pi}{v_F} \int_{\vec{k}_1,\vec{q}} \delta(k-k_1-|\vec{k}-\vec{q}|+|\vec{k}_1+\vec{q}|) R_1(\vec{k},\vec{k}_1,\vec{q})  \nonumber \\ &&  \left\{ f_{\lambda}(\vec{k}) f_{-\lambda}(\vec{k}_1) [1-f_{\lambda}(\vec{k}-\vec{q})][1-f_{-\lambda}(\vec{k}_1+\vec{q})]  - [1-f_{\lambda}(\vec{k})] [1-f_{-\lambda}(\vec{k}_1)]f_{\lambda}(\vec{k}-\vec{q}) f_{-\lambda}(\vec{k}_1+\vec{q})\right\} \nonumber \\ && 
-\frac{2\pi}{v_F} \int_{\vec{k}_1,\vec{q}}\delta(k+k_1-|\vec{k}-\vec{q}|-|\vec{k}_1+\vec{q}|) R_2(\vec{k},\vec{k}_1,\vec{q}) \nonumber \\ &&  \left\{ f_{\lambda}(\vec{k}) f_{\lambda}(\vec{k}_1) [1-f_{\lambda}(\vec{k}-\vec{q})][1-f_{\lambda}(\vec{k}_1+\vec{q})]  - [1-f_{\lambda}(\vec{k})] [1-f_{\lambda}(\vec{k}_1)]f_{\lambda}(\vec{k}-\vec{q}) f_{\lambda}(\vec{k}_1+\vec{q})\right\}, \nonumber \\
\end{eqnarray}
with

\begin{eqnarray}
R_1(\vec{k},\vec{k}_1,\vec{q}) &=&  4\left( \left| T_{++--}(\vec{k},\vec{k}_1,\vec{q}) - T_{+-+-}(\vec{k},\vec{k}_1,\vec{k}-\vec{q}-\vec{k}_1)\right|^2 \right. \nonumber \\ &&\left. + (N-1)\left[ \left| T_{++--}(\vec{k},\vec{k}_1,\vec{q})\right|^2+ \left| T_{+-+-}(\vec{k},\vec{k}_1,\vec{k}-\vec{q}-\vec{k}_1)\right|^2 \right] \right),\nonumber \\
R_2(\vec{k},\vec{k}_1,\vec{q}) &=&  4\left(\frac{1}{2} \left| T_{++++}(\vec{k},\vec{k}_1,\vec{q}) - T_{++++}(\vec{k},\vec{k}_1,\vec{k}-\vec{q}-\vec{k}_1)\right|^2  + (N-1) \left| T_{++++}(\vec{k},\vec{k}_1,\vec{q})\right|^2 \right).\nonumber\\
\end{eqnarray}
 Here we introduce the short-hand:
\begin{equation}
T_{\lambda_1\lambda_2\lambda_3\lambda_4}(\vec{k},\vec{k}_1,\vec{q}) = \frac{V(\vec{q})}{2} \left( \mathcal{M}^{\lambda_1\lambda_2}_{\vec{k},\vec{k}-\vec{q}} \mathcal{M}^{\lambda_3\lambda_4}_{\vec{k}_1,\vec{k}_1+\vec{q}}\right) ,
\end{equation}
with the form factor:
\begin{equation}
M^{\lambda\lambda_1}_{\vec{k},\vec{k}_1} = \left[ \mathcal{U}^\dagger_{\vec{k}} \mathcal{U}_{{\vec{k}_1}}  \right]_{\lambda\lambda_1} =   \frac{1}{2} \left(1 + \lambda \lambda_1 e^{i\left( \theta_{\vec{k}}-\theta_{\vec{k}_1}\right)}\right).             
\end{equation}
Here $\theta_{\vec{k}} = \atan(k_x/k_y)$ is the angle of the momentum $\vec{k}$ with respect to the $x$-axis. The Coulomb potential energy in the Fourier space reads as $V(\vec{q}) = 2\pi \alpha v_F /q$. Note that this collision integral can alternatively be determined by application of Fermi Golden's rule. 

The above Boltzmann equation was solved by the variational method \cite{Ziman2001} for shear viscosity of graphene at half-filling \cite{Fritz2009}. The same approach will be used here to calculate the current diffusion coefficient $\zeta$. To this end, we first linearize the Boltzmann equation by the ansatz
\begin{equation}
	f_{\lambda} = \tilde{f}_{\lambda}^0 + \delta f_{\lambda}.
\end{equation} In the presence of an external electric field, electrons and holes at CNP counter-propagate with opposite fluid velocity $\lambda \vec{u}$. Each of them forms individual fluids with the equilibrium distribution $ 
	\tilde{f}^0_\lambda = \frac{1}{e^{(\epsilon_\lambda(\vec{k}) -\lambda \vec{u}\cdot\vec{k})/T}+1}$.  
As a result, total charge current is given by  
\begin{equation}
\vec{j} = N e\sum_{\lambda=\pm}\int_{\vec{k}} \vec{v}_{\lambda} \tilde{f}^0_\lambda(\vec{k}) = en_{\text{th}} \vec{u} +\mathcal{O}(u^2),
\end{equation}
with $N = 4$ accounting for valley and spin degrees-of-freedom and the number density of thermal excitations $n_\text{th} =  \frac{4\pi T^2}{12 v_F^2}$ remaining non-zero even at charge neutrality. In linear response,
the deviation of the distribution function from the equilibrium $\delta f_\lambda$ is linear in the shear force 
\begin{equation}X^{\lambda}_{ij} = \frac{\lambda}{2} \left( \partial_{i}u_j + \partial_ju_i - \delta_{ij}\vec{\nabla}\cdot \vec{u}\right).
\end{equation}
This force is applied to the system via a current gradient $\partial_i j_j$ which amounts to the fluid velocity gradient $\lambda \partial_i u_j$. We write
\begin{equation}
\delta f_\lambda = \frac{g(\frac{\epsilon_\lambda(\vec{k})}{T})}{T} \frac{\lambda v_F k}{T} \left(\frac{k_ik_j}{k^2} - \delta_{ij} \right)  f_\lambda^0(\vec{k}) \left( 1- f_\lambda^0(\vec{k})\right)  X^\lambda_{ij},
\end{equation}
with the dimensionless and unknown function  $g(\frac{\epsilon_{\lambda}(\vec{k})}{T})$ to be determined by solving the linearized Boltzmann equation.

It is worth pointing out, as a comparison, that, in the calculation of  shear viscosity \cite{PhysRevLett.103.025301,Kitinan2024}, electrons and holes form a single fluid and thus propagate with a joint fluid velocity $\vec{u}$. The equilibrium solution is instead given by $\tilde{f}^0_\lambda = \left(e^{(\epsilon(\vec{k}) - \vec{u}\cdot\vec{k})/T}+1\right)^{-1}$ and the shear force in this case is given by $X_{ij} = \frac{1}{2} \left( \partial_{i}u_j + \partial_ju_i - \delta_{ij}\vec{\nabla}\cdot \vec{u}\right)$.

We determine the current diffusion coefficient $\zeta$ by computing the current flux: 
\begin{eqnarray}
\label{eq:currentflux}
{{\mathcal{J}}}_{mn} &=&N e\sum_{\lambda} \int_{\vec{k}} \vec{v}_{\lambda m}\vec{v}_{\lambda n} \delta f_\lambda \nonumber\\
&=& Ne\sum_{\lambda} \int_{\vec{k}} \vec{v}_{\lambda m}\vec{v}_{\lambda n} \frac{g(\frac{\epsilon_\lambda(\vec{k})}{T})}{T} \frac{\lambda v_F k}{T} \left(\frac{\vec{k}_r\vec{k}_s}{k^2} - \delta_{rs} \right) X^\lambda_{rs}  f_\lambda^0(\vec{k}) \left( 1- f_\lambda^0(\vec{k})\right)
\end{eqnarray}
The current diffusion coefficient can be extracted from a constitutive relation:

\begin{equation}
{{\mathcal{J}}}_{xy} = \zeta {\nabla}_x {j}_y, 
\end{equation}
with
\begin{equation}
\zeta = \frac{1}{4n_{\text{th}}}\sum_{\lambda} \int_{\vec{k}}f_\lambda^0(\vec{k}) \left( 1- f_\lambda^0(\vec{k})\right) \frac{v_F^3k}{T^2}g\left(\frac{\epsilon_\lambda(\vec{k})}{T}\right).
\end{equation}

Solving the linearized Boltzmann equation for $g\left(\frac{\epsilon_\lambda(\vec{k})}{T}\right)$ by the variational method \cite{Fritz_2008,Fritz2009,Ziman2001}, we find the current diffusion coefficient of graphene resulting as 
\begin{equation}
\frac{\zeta}{a^2t} = C \left( \epsilon a t\right)^2 \frac{t}{T},
\end{equation}
with $C=25.76$, and dimensions written explicitly using C-C hopping $t$ as a unit of energy and graphene unit cell length $a$ as a unit of distance.

 \begin{figure}[]
  \centering
\includegraphics[width=0.49\textwidth , angle=0]{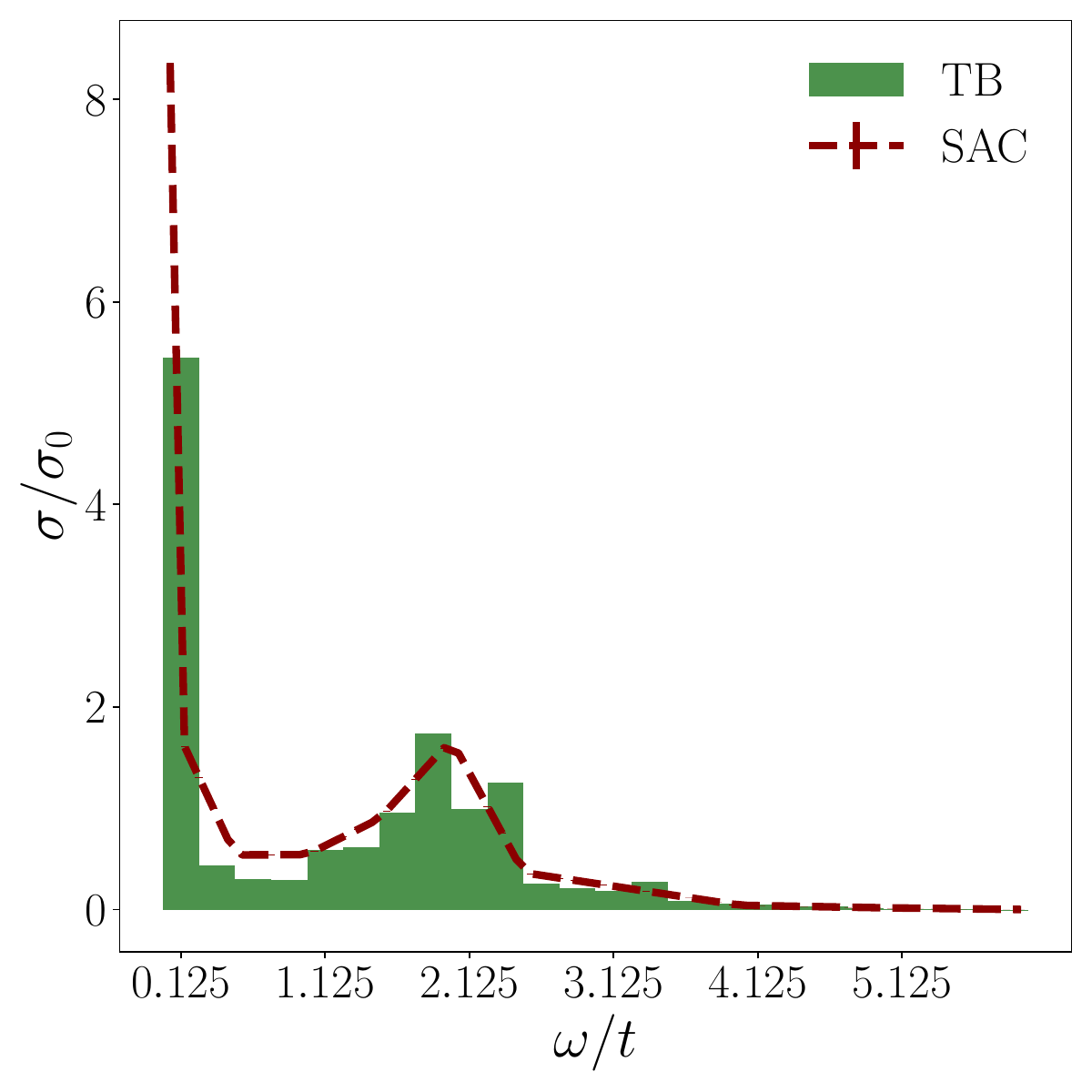}
        \caption{Comparison of the analytical conductivity for free tight-binding model and SAC results obtained from corresponding free current-current correlator. Calculations were made for the graphene strip with 27 stripes, $\beta t=1.5$. Spectral function corresponds to $\sigma_{yy}(x, \omega)|_{x=13}$. }
  \label{fig:SAC_TB_test}
\end{figure}

\section{Computation of conductivity and the role of adatoms}

\subsection{Calculating the conductivity with QMC and the Stochastic analytical continuation method}

 Generally speaking, real-time dependent observables are difficult to measure with Quantum Monte Carlo (QMC) \cite{SHAO20231}. At  equilibrium  we have to rely on the analytical continuation \cite{Sandvik98,Beach04a,SHAO20231} of the Euclidean current-current correlator, describing the reaction of the current density in a stripe defined by its coordinate $x$ to the uniform electrical field in $y$ direction applied to the whole sample:
\begin{equation}
G_{yy}(x, \tau) = \sum_{ \substack{\left< i,j\right > \in x \ \\ \left< m,n\right >}} T^{ij}_{y} T^{mn}_{y}  \mathrm{Tr}  \left ( \hat J_{i,j}e^{-\tau \hat H} \hat J_{m,n} e^{-(\beta-\tau) \hat H} \right ).
\label{eq:sm:correlator}
\end{equation}
The current operator is defined as $\hat J_{i,j} = i t \left(\hat a_{i, \sigma}^{\dag} \hat a_{j, \sigma} - \hat a_{j, \sigma}^{\dag} \hat a_{i, \sigma} \right)$. The normalization factor $ T^{ij}_{y} = \Vec{e}_{ij} \cdot \Vec{e}_y$ takes into account the geometry of the lattice and direction of the bond $\Vec{e}_{ij}$ between sites $i$ and $j$ with respect to the electric field. The correlator \eqref{eq:sm:correlator} is computed using determinantal Quantum Monte Carlo \cite{Blankenbecler81,White89,Assaad08_rev}. We employ the ALF package \cite{ALF_v2}, using the finite temperature QMC implementation with continuous fields with a Trotter decomposition step $\Delta \tau = 0.05$ in units of the inverse hopping $t^{-1}$.  A typical sample has 12 lattice steps in the direction of periodical boundary conditions (see Fig. 1 in the main text) and varying width in the direction of open boundary conditions. Disorder is introduced along the open edges by randomly placing adatoms.  We generate  between 20 and 320 different configurations of adatoms (number depending on the lattice size) for each temperature and lattice, perform QMC simulations separately for each configuration of adatoms and subsequently average the results over them. The width of the samples is increasing with decreasing temperatures. A list of temperatures and corresponding lattice sizes can be found in Tab.~\ref{tab:latticesizes}. 

\begin{table}[h]
\centering
\begin{tabular}{|l|l|l|l|l|l|l|l|l|} 
\hline
Stripes $N_s$                     & 18  & 21   & 24   & 27   & 30   & 34   & 36   & 45    \\ 
\hline
$\beta t$ & 1.0 & 1.15 & 1.3  & 1.5  & 1.65 & 1.8  & 2.0  & 2.5   \\ 
\hline
$T/t$                   & 1.0 & 0.87 & 0.77 & 0.67 & 0.60 & 0.56 & 0.50 & 0.40  \\ 
\hline
Samples                    & 15  & 30   & 30   & 40   & 60   & 80   & 120  & 320   \\
\hline
\end{tabular}
\caption{Simulated lattice sizes with corresponding temperatures and number of samples. In order to generate enough statistics to resolve the profiles in the lattice, we used a large number of samples for the larger lattices (lower temperatures).}
\label{tab:latticesizes}
\end{table}

 The low frequency (hydrodynamic regime \cite{Fritz_2008}) conductivity $\sigma_{yy} (x,\omega = 0)$ can be subsequently extracted from the correlator \eqref{eq:sm:correlator} using the stochastic analytical continuation method (SAC) \cite{Sandvik98,Beach04a,SHAO20231} exploiting the Green-Kubo relations:
\begin{equation}
G_{yy}(x,\tau) = \int_0^{\infty} K(\omega, \tau) \, \sigma_{yy}(x,\omega) \, d\omega
\label{eq:GreenKubo}
\end{equation}
with the thermal kernel $K(\omega, \tau) = \frac{2\omega \, \cosh \left( \omega \left( \tau - \frac{1}{2T} \right) \right)}{\sinh \left( \frac{\omega}{2T} \right)}$.  In order to check the ability of SAC to reproduce the qualitative features of the frequency-dependent conductivity, we performed the following test: we took a set of 20 samples with width $N_s=27$ and temperature $\beta t=1.5$ (one of the setups used for QMC), and computed the conductivity for free tight-binding model both analytically and using SAC starting from corresponding free current-current correlator. Results are shown in the Fig.~\ref{fig:SAC_TB_test}. As one can see, despite the high temperature, SAC can reproduce all qualitative features of the analytical spectral function: Drude peak at small $\omega \rightarrow 0$, drop-down at intermediate frequencies, and the effect of the van Hove singularity at $\omega=2$. Thus we can conclude that SAC can still give us reliable qualitative results.

The resulting data for the DC conductivity, shown in Fig.~\ref{fig:QMCandMEMProfiles_b} is then obtained by applying SAC to the QMC data of one given adatom configuration and consequently averaging over all configurations. The errorbars correspond to the error of the average over adatom configurations and the data is also symmetrized with respect to the middle stripe of the sample.

  \begin{figure}[]
   \centering
   \subfigure[]     { \label{fig:QMCandMEMProfiles_a}\includegraphics[width=0.49\textwidth]{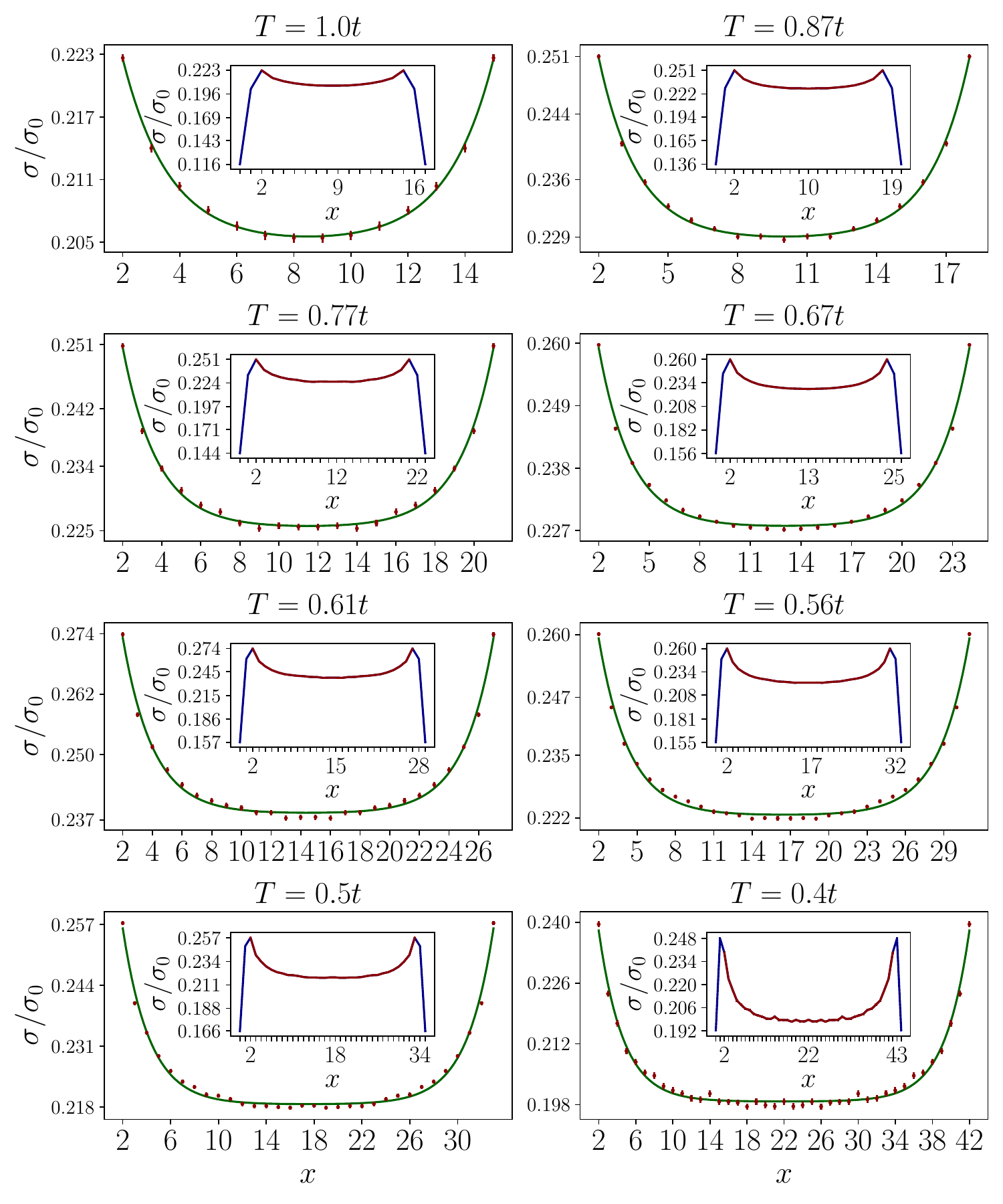}}
   \subfigure[]    { \label{fig:QMCandMEMProfiles_b}\includegraphics[width=0.49\textwidth]{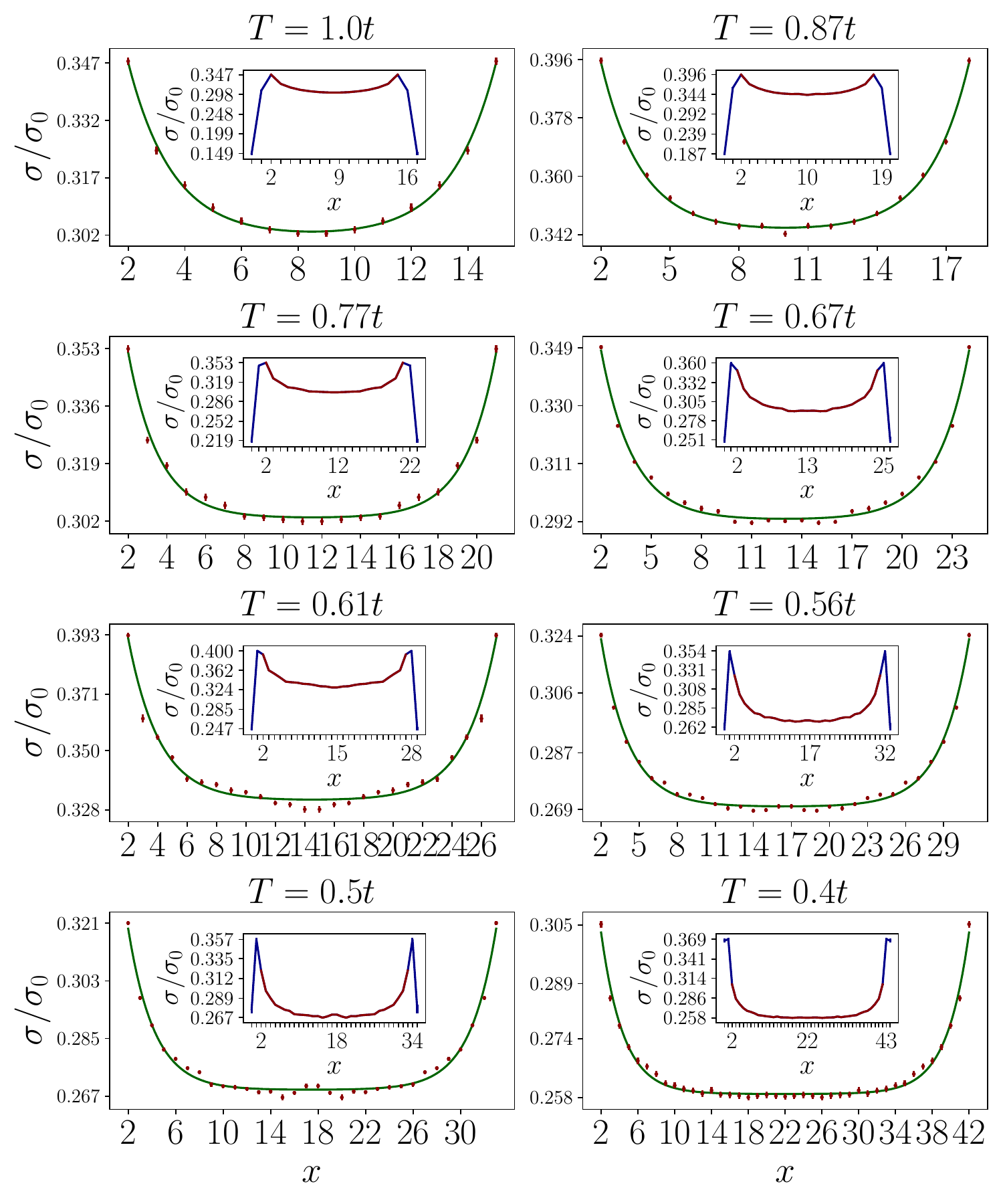}}
         \caption{Fitted conductivity profiles from the middle point of the correlator (a) and SAC (b) after averaging over all adatom configurations and symmetrizing around the middle of the sample. The raw data is shown as an inset for each profile, where the red curve corresponds to the data used for the fit (we start from the first point after the maximum). In the SAC case, the errobars of each profile correspond to the error of the average over all adatom configurations. The ftting function (green) is the catenary curve from Eq.~(6) in the main text. }
    \label{fig:QMCandMEMProfiles}
 \end{figure}

 The low frequency conductivity can be also approximated by using the middle point of the current-current correlator \eqref{eq:sm:correlator} \cite{Trivedi1996,Buividovich12,PhysRevB.94.085421, PhysRevLett.118.266801}:
\begin{equation}
\sigma(x) \approx \frac{1}{\pi T^2} \int_0^\infty \frac{d\omega}{2\pi} \frac{2 \omega \sigma(\omega)}{\sinh \left( \frac{\omega}{2T} \right)}  = \frac{1}{\pi T^2}G_{yy}(x,\beta/2)
\label{eq:sm:middle_point}
\end{equation}
These profiles are shown in Fig.~\ref{fig:QMCandMEMProfiles_a}. They are again obtained from averaging over the QMC data of all adatom configurations and symmetrizing around the middle-stripe.

  \begin{figure}[]
   \centering
   \subfigure[]     {\label{fig:T_dep_corr_a}\includegraphics[width=0.49\textwidth]{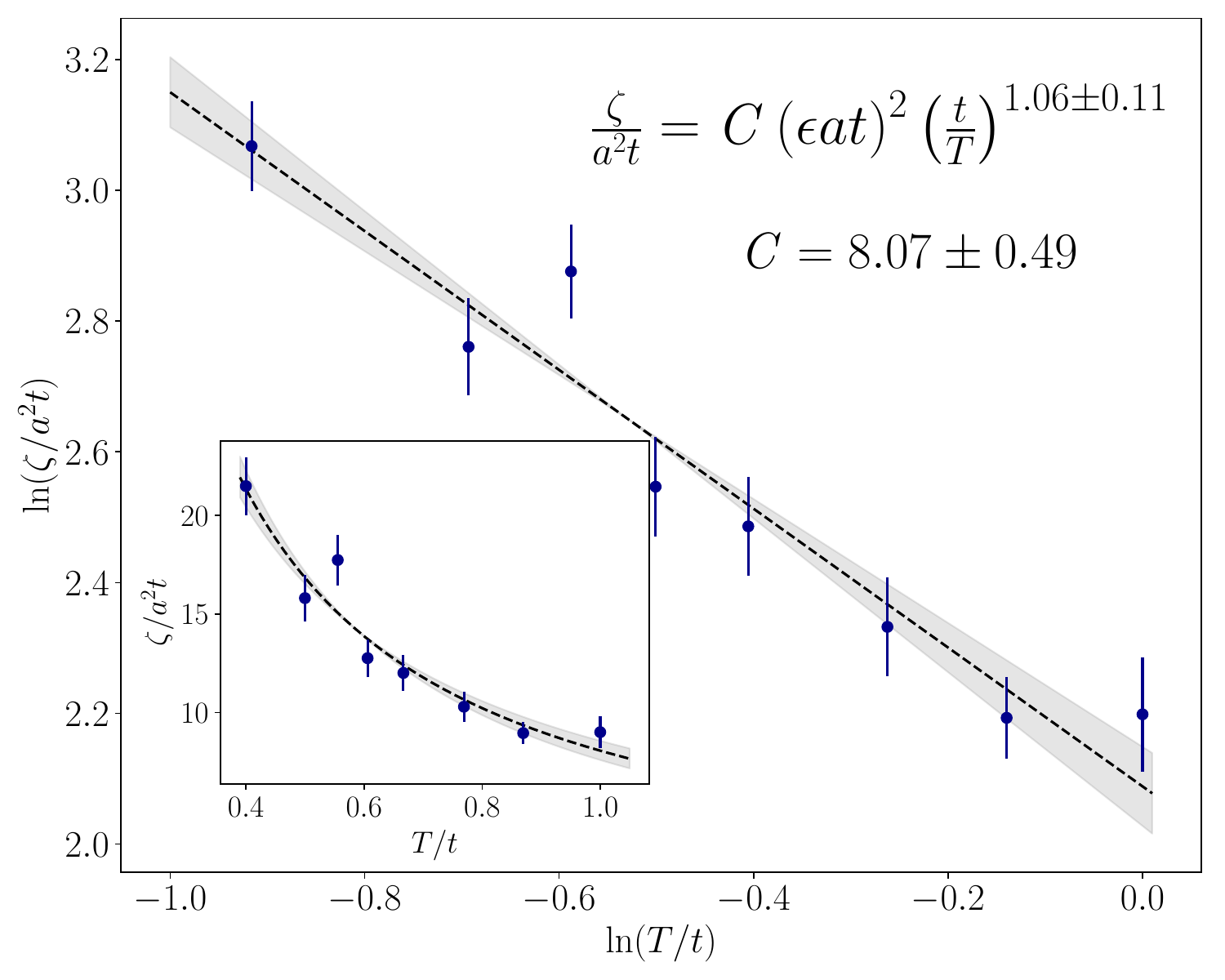}}
   \subfigure[]    {\label{fig:T_dep_corr_b}\includegraphics[width=0.49\textwidth]{figures/ViscosityWithFit_InsetMu.pdf}}
          \caption{ Temperature dependent current diffusion coefficent $\zeta$ in log-log representation (main plots) and with linear scales (insets). The power law fit is shown with the dotted line. (a) Data is obtained from the conductivity profiles computed using the middle point of the current correlator  \eqref{eq:sm:middle_point}. (b) Data us obtained using the full SAC spectral functions. }
\label{fig:T_dep_corr}
 \end{figure}

 Comparing the conductivity profiles calculated from the middle point of the correlator and SAC, we notice that they all agree well with the Eq.~(6) in the main text. We observe finite size effects, such as small deviations from the smooth catenary curve (in particular for smaller temperatures), which also show up in small deviations from the power law in both SAC and correlator middle-point data (see Fig.~\ref{fig:T_dep_corr}). Nevertheless, as apparent from Fig.~\ref{fig:T_dep_corr}, the $1/T$ temperature dependence of the resulting current diffusion coefficient is evident in both cases. The similarities between the correlator middle-point and SAC profiles are also another marker for the smoothness of the frequency dependent conductivity $\sigma(\omega)$ at small $\omega$, which is also visible in Fig.~\ref{fig:SigmaOmega}\textcolor{red}{(d)}. The remaining difference between correlator middle-point and SAC data (note the different values of the C coefficient in Fig.~\ref{fig:T_dep_corr} ) can be attributed to the fact, that the resolution function $\frac{2 \omega}{\sinh \left( \frac{\omega}{2T} \right)}$ in \eqref{eq:sm:middle_point} has in fact a finite width $\simeq T$, hence the approximate conductivity does not exactly correspond to the DC one.  Despite these deficiencies, the usage of the middle point correlator can still be beneficial to recover the qualitative behavior of the low-frequency part of the spectral functions, because we can avoid larger fluctuations caused by the inherent instability of the analytical continuation techniques. 

 Summarizing, the qualitative behavior of the conductivity profiles across the sample and the temperature dependence of the current diffusion coefficient are independent on the underlying method for the conversion of the Euclidean current-current correlator into the low-frequency conductivity. It justifies the smoothness of the spectral functions at $\omega \rightarrow 0$ and supports the claim that our SAC results shown in the main text are valid despite the instability of the analytical continuation. 

 Next, we discuss in more detail the role of disorder imposed by adatoms on the formation of the boundary conditions.

\subsection{Influence of adatoms on the conductivity}
 In order to quantify the effects caused by adatoms, we first look at the density of states (DoS) for a free tight-binding Hamiltonian (Fig.~\ref{fig:DOS}). Adatoms create localized near-zero and high-energy modes. They correspond to energies around $E =0$ in the DoS on Fig. \ref{fig:DOS} and around $E=10.0$. Bulk energy levels have energies $ t > |E| > 0$. The near-zero modes are located on the sites next to the ones where the adatom is attached, whereas the high-energy states belong to the adatoms themselves. These near-zero modes might be actually responsible for the inverse boundary conditions we already observed for the optical conductivity in the free case. 

 In order to understand this phenomenologically, we computed the frequency-dependent free conductivity (see Fig.~\ref{fig:SigmaOmega}\textcolor{red}{(a)} and \textcolor{red}{(b)}) for a lattice of 27 stripes (i.e. 648 orbitals) and the disorder configuration with 8 (2) adatoms on the edge stripes. Figure \ref{fig:SigmaOmega}\textcolor{red}{(a)} reveals a strong dependence of the profile on the frequency $\omega$. The bulk conductivity is maximal at low frequencies $\omega \leq 0.2t$. This corresponds to the Drude peak (see Fig.~\ref{fig:SigmaOmega}\textcolor{red}{(b)}) which contributes significantly to the midpoint of the current correlator. The midpoint approximation \eqref{eq:sm:middle_point} is therefore no longer valid and we expect inadequate results for an ideal non-interacting system. At larger frequencies around the Dirac-plateau regime $0.2t < \omega < 1.5t $, the profile is inverted, and the bulk conductivity reaches its minimal value whereas the conductivity at the edges is bigger.

 \begin{figure}[h]
  \centering
\includegraphics[width=0.45\textwidth , angle=0]{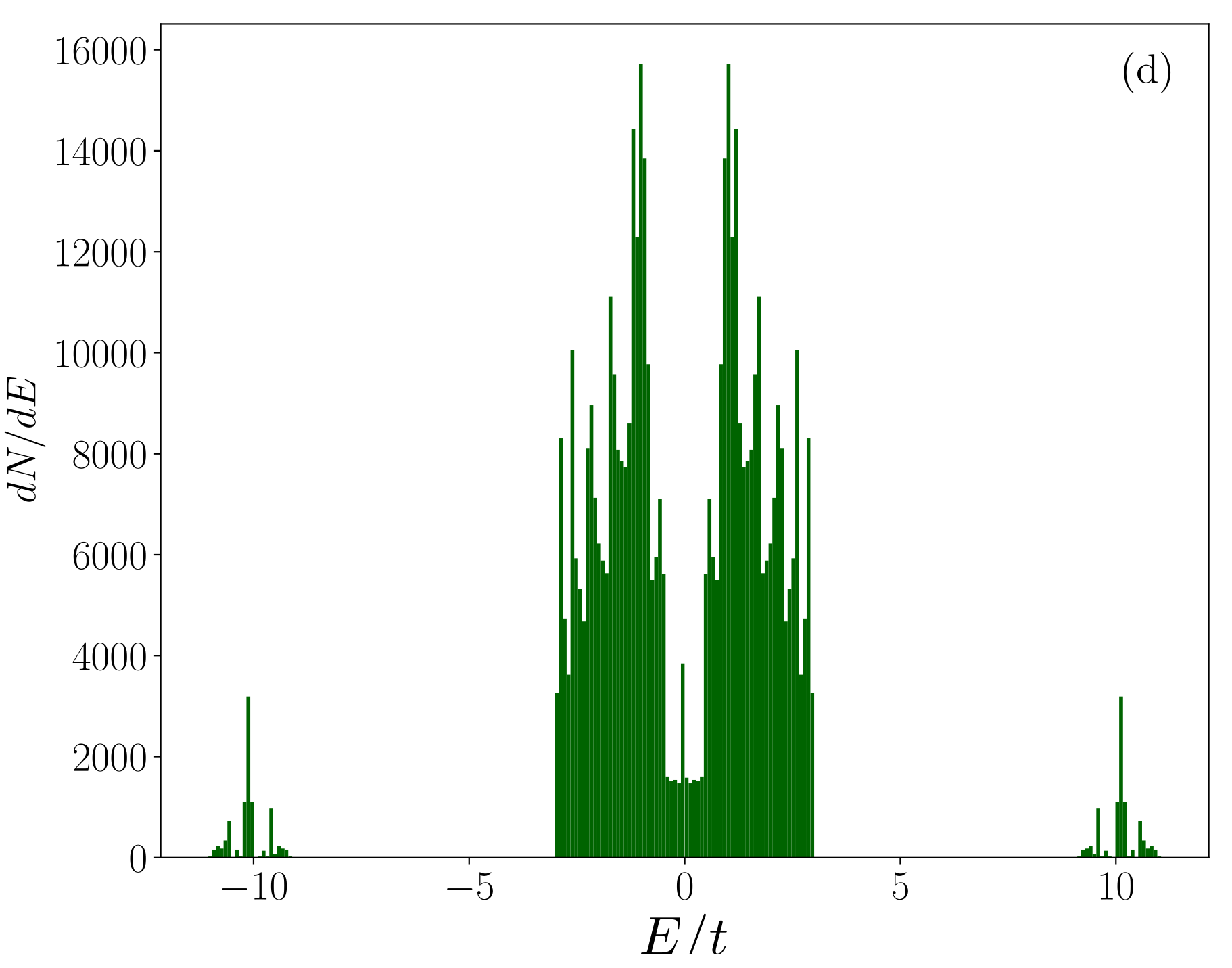}
\caption{Density of states for the samples with width $L_2 = 90$ (45 stripes). We averaged the data over 20 samples with different configurations of adatoms at the edges to get a better energy resolution.}
\label{fig:DOS}
\end{figure}

 \begin{figure}[h]
  \centering
\includegraphics[width=0.9\textwidth , angle=0]{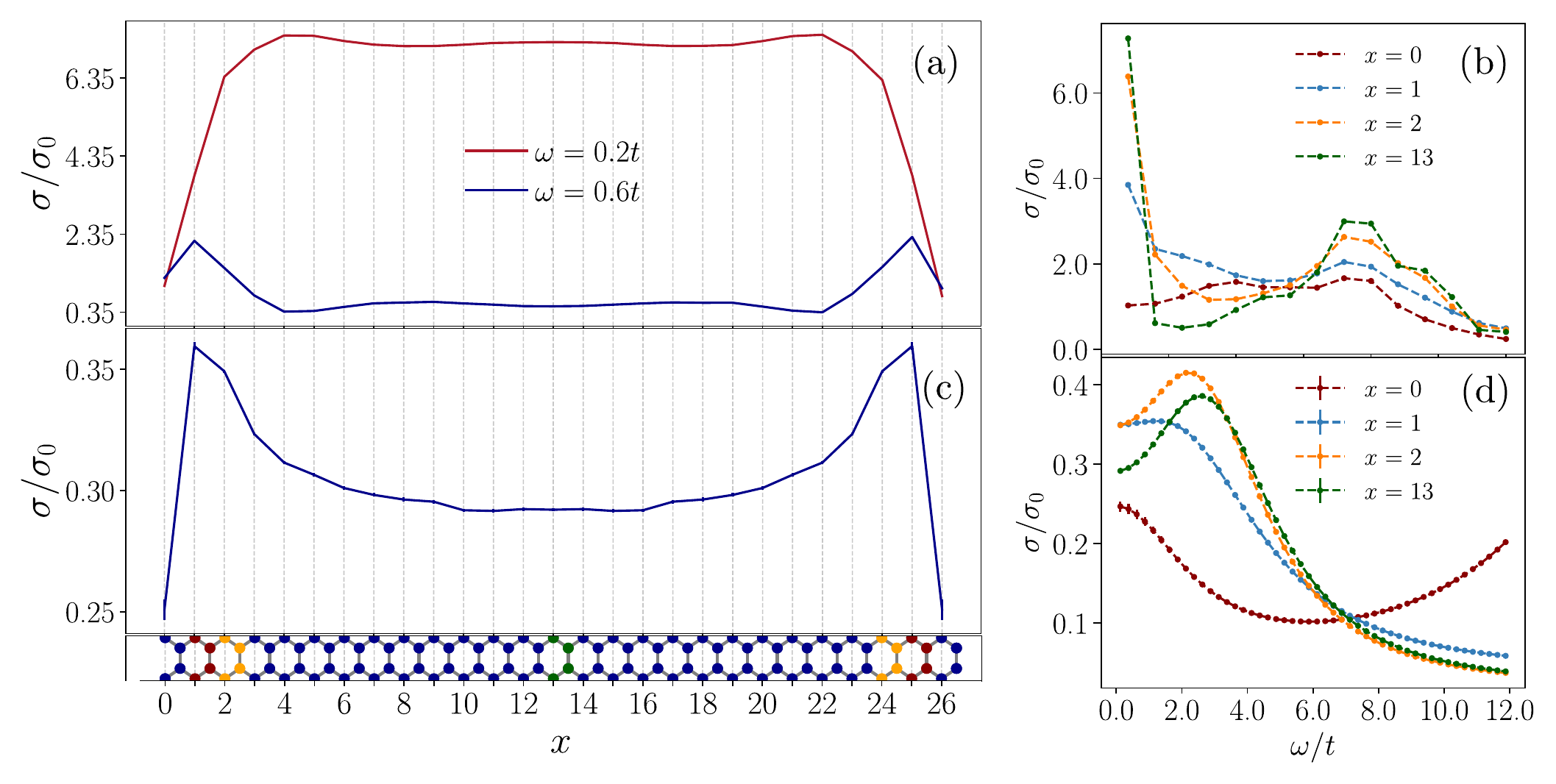}
\caption{(a) Local conductivity profiles at two different frequencies for the free tight-binding model with randomly distributed adatoms at the edges at a temperature $T=0.67t$. (b) shows $\sigma(\omega)$ profiles for the same free Hamiltonian at different stripes of the lattice. (c) Local DC conductivity profile across the sample in the interacting case for the same lattice and temperature. SAC data, obtained from the Euclidean current-current correlator.(d) is showing the frequency dependent conductivities of the same stripes as in (b) but for this interacting system. Below the profile figures, we sketched the corresponding lattice width, with the edge (red), boundary layer (orange) and bulk (green) stripes colored.}
\label{fig:SigmaOmega}
\end{figure}

 To verify that the midpoint of the current-current correlator is a valid estimator in the interacting case, we extract the conductivity using SAC from the QMC data of the above system size and compare it to the free case. The results in Fig.~\ref{fig:SigmaOmega}\textcolor{red}{(d)} 
show that the Drude peak is indeed suppressed for the bulk conductivity, thus yielding a smaller value than at the boundary for small frequencies. The boundary conditions have therefore the same effect on the conductivity profile as in the non-interacting regime with frequencies around the Dirac plateau, and the midpoint of the correlator is a valid estimator for the DC conductivity.

 We should stress that our analysis is valid for suspended graphene. In typical hydrodynamic experiments however, graphene samples are encapsulated between two layers of hexagonal boron nitride\cite{Ku2020}. The Coulomb interaction is therefore weaker and the reduction of the Drude peak is less pronounced. As a consequence, the boundary conditions need to be reevaluated depending on the interaction strength. Also, with a weaker interaction, the approximation using the midpoint of the current-current correlator can become erroneous. However, we expect that the SAC processing remains stable at all interaction strengths. 

 \begin{figure}[]
  \centering
\includegraphics[width=0.49\textwidth , angle=0]{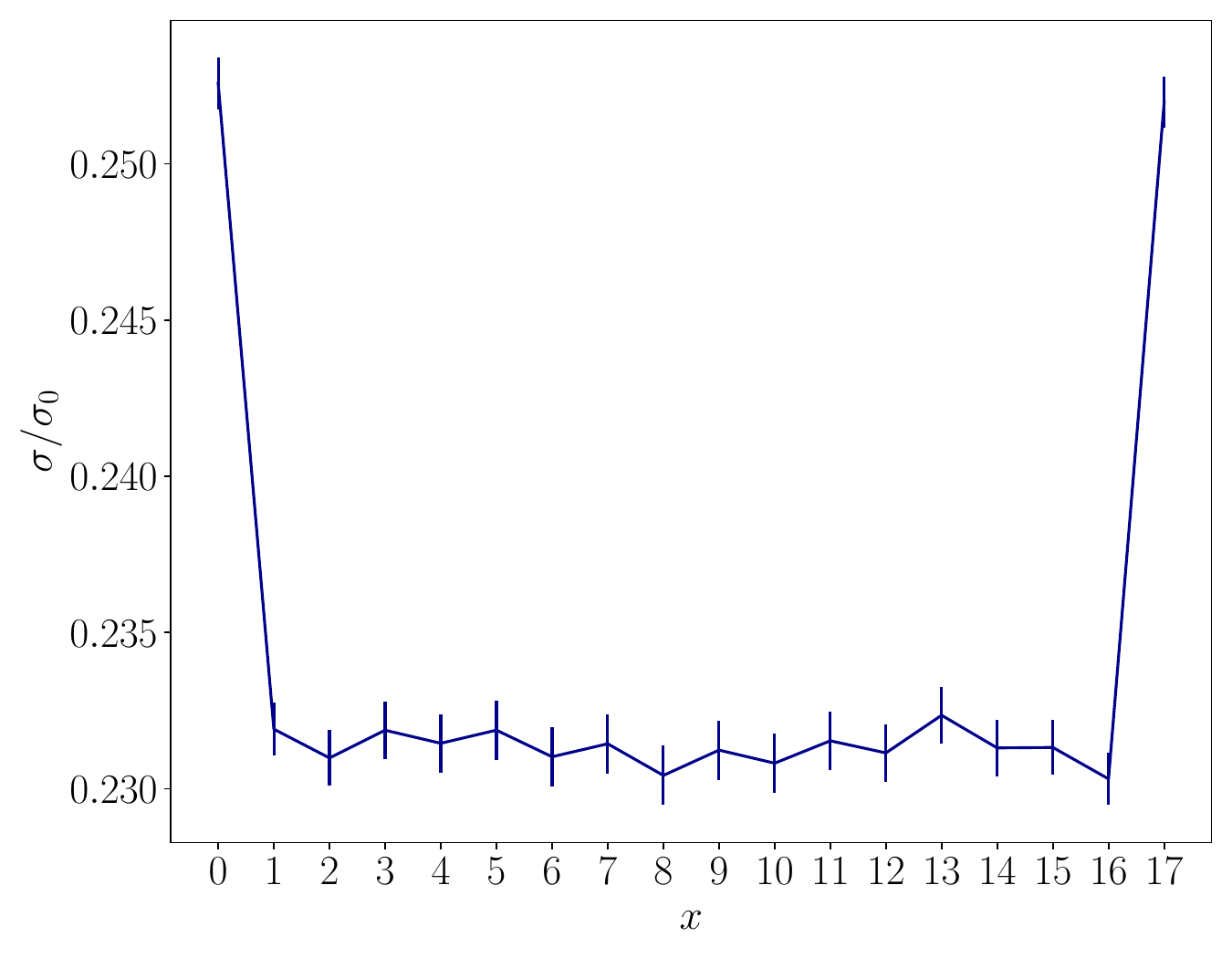}
        \caption{Flat QMC conductivity profile of a lattice with open zigzag edges due to a low scattering rate between edge and bulk wavefunctions. Calculations are done at $T=t$ for the standard interaction strength corresponding to suspended graphene.}
  \label{fig:ZigZagProfile}
\end{figure}

On a microscopic level, the disorder at the edges is essential for creating some sort of boundary conditions, since we need non-momentum conserving scattering of electronic states from the edges to create a non-uniform profile of charge current or momentum. 
This effect can be also checked numerically. We can consider other types of boundaries, in particular zigzag ones, which also have inherent zero modes, albeit without any disorder. Results of the full QMC calculation for the clean zigzag edge are shown in the Fig. \ref{fig:ZigZagProfile}. As one can see, the zero modes at the border create an exceptional point right at the edges, but we do not see any hydrodynamic profile due to y-component of momentum being conserved even after scattering of particles from the edge.

\section{Numerical evaluation of ${\mathcal{E}_{\pm}^{mn}}$}

 For the computation of the current diffusion coefficient from the numerical data, it is necessary to calculate the diagonal element of ${\mathcal{E}_{\pm}^{mn}}= e\int d{\vec{k}} \partial_{m}{v}^{n}_{\pm} \tilde{f}_{\pm}$ (see main text) for the simulated lattices. In order to evaluate the integral over the discrete Brillouin-zone of our finite-sized lattices, we use the exact interacting Green Functions $G(\vec{k},\nu_1,\nu_2,\sigma) = \left< \hat{c}_{\vec{k}\sigma\nu_1}^\dagger \hat{c}_{\vec{k}\sigma\nu_2} \right>$ obtained in QMC for the evaluation of the distribution functions $\tilde{f}_{\pm}(\vec{k})$, where $\nu_1 $and $\nu_2$ correspond to the sublattice indices. More precisely we have 
\begin{equation}
\tilde{f}_{\pm}(\vec{k}) =  \left< \hat{\Psi}_{\pm}^\dagger (\vec{k}) \hat{\Psi}_{\pm} (\vec{k}) \right>=\sum_{\sigma,\nu_1,\nu_2} \psi_{\pm,\nu_1,\sigma}^* \psi_{\pm,\nu_2,\sigma}\left<  \hat{c}_{\vec{k}\sigma\nu_1}^\dagger \hat{c}_{\vec{k}\sigma\nu_2} \right>
\end{equation}
where $\psi_{\pm,\nu_1,\sigma}$ are the exact wavefunctions of holes (+) and electrons (-). obtained from the free tight-binding Hamiltonian.

 For the evaluation of  $\partial_{m}{v}^{n}_{\pm} =  \partial_{m}\partial_{n}E_{\pm}(\vec{k})$ we use the free dispersion relation $E_{\pm}(\vec{k})$ of graphene, justified by only minor changes in the dispersion relation for the interacting case for the energies away of the Dirac point \cite{PhysRevB.110.155120}.
Therefore, the discretized integral for $\mathcal{E}_{\pm}^{mn}$ becomes 
\begin{equation}
{\mathcal{E}_{\pm}^{mn}}= \sum_{\vec{k} \in BZ} \partial_{m}\partial_{n}E_{\pm}(\vec{k}) \cdot \tilde{f}_{\pm}(\vec{k}) \cdot \Delta_{\vec{k}}
\label{eq:DiscreteEpsilon}
\end{equation}
where the summation is carried out over the finite set of quantized $\vec{k}$ points in the Brillouin-zone of the lattice and $\Delta_{\vec{k}} = \left| \vec{g_1}\cross\vec{g_2}\right|$ corresponds to the area element in the discretized $\vec{k}$-space. The unit vectors $\vec{g_1}$ and $\vec{g_2}$ are defined such that each discrete point $\vec{k}$ in the Brillouin-zone can be accessed via $\vec{k} = n\vec{g_1} + m\vec{g_2}$ with $n,m \in \mathbb{Z}$.

 In Fig.~\ref{fig:CurrentDiffandEpsilon} we compare the values of $\mathcal{E}_{\pm}^{nn} \equiv \mathcal{E}$ computed using the exact integral expression for the free case with the free tight-binding model computation on a finite sample, and with the full QMC calculation. We notice a quantitative difference of approximately $15\%$ between the exact and tight-binding data even for the free system. This comes from the evaluation of $ \mathcal{E}$ on a finite-sized lattice. Thus the increase of lattice size for a given temperature should yield a better estimate. The inset (in Fig.~\ref{fig:CurrentDiffandEpsilon}) shows this, where the TB data points for $\mathcal{E}$ at fixed temperature $T = t$ approach the exact integral with the increase of lattice size $L$. The obvious reason is that with the increase of the number of $\vec{k}$-points the evaluation of the integral via the sum becomes more accurate and approaches the exact value for large enough lattice sizes. In general, we can conclude that the qualitative behavior of $ \mathcal{E}$ for our systems is in accordance with the exact result and only deviates by $15\%$. An additional observation is that at least for $ \mathcal{E}$, our temperatures are still far from the Dirac point, where we expect  $ \mathcal{E}/T$ to converge to $-2\ln2/\pi$. Finally, the figure also demonstrates that interaction effects become noticeable for $\mathcal{E}$ only when approaching temperatures below $0.6t$. Despite being marginal, these effects have to be taken into account for a better estimation of $\zeta$.

\begin{figure}[h]
        \centering
        \includegraphics[width=0.5\linewidth]{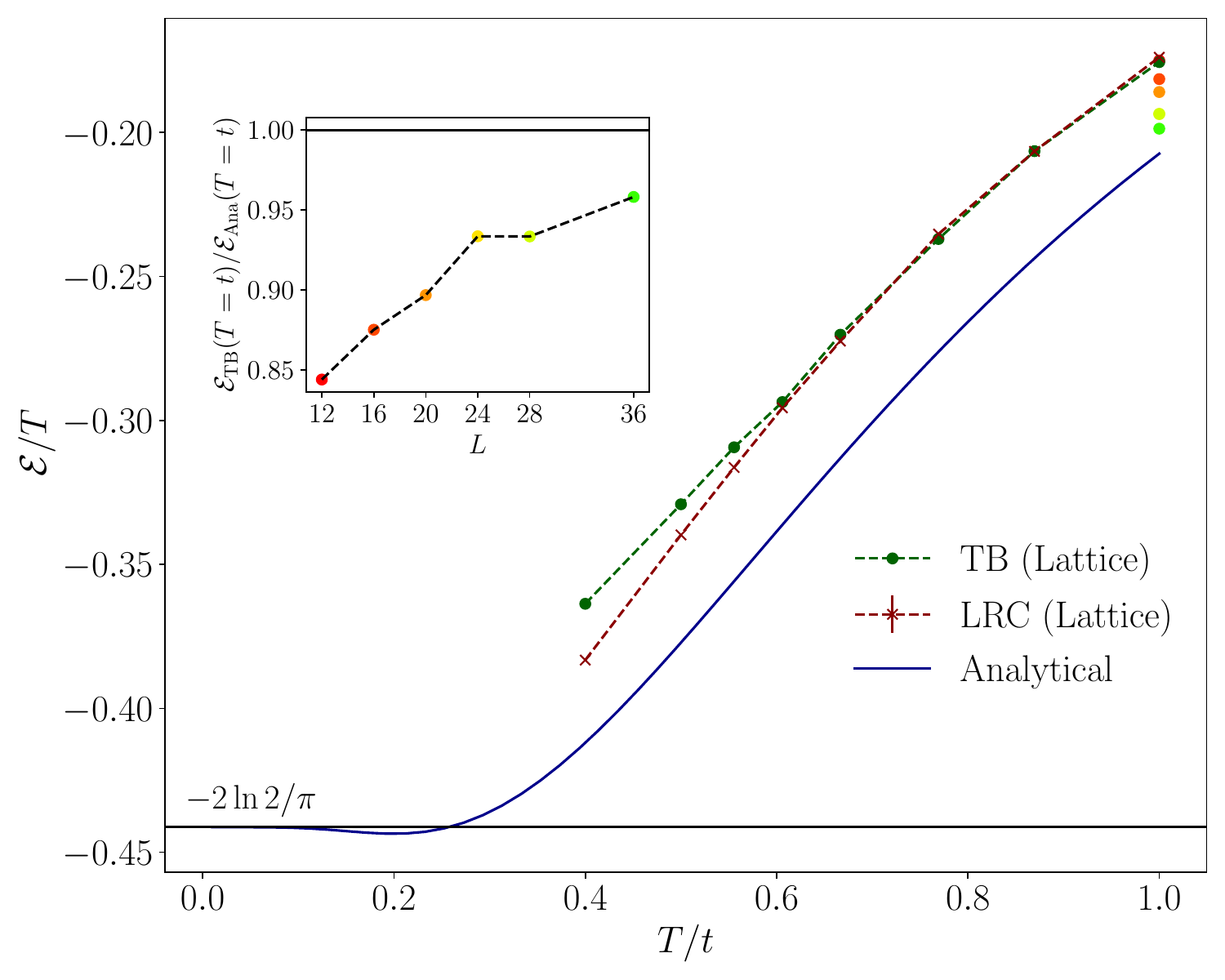}
    \caption{Computation of $\mathcal{E}$. For the free case, we compare tight-binding calculation with the exact integral. Inset shows the convergence of the tight-binding $\mathcal{E}$ towards the exact integral with increasing lattice sizes at $T = 1.0t$. Main plot shows both the free and the interacting (LRC) case.}
    \label{fig:CurrentDiffandEpsilon}
\end{figure}

\section{Controlling the transport regime}
 Within the framework of our microscopic description of graphene, we can distinguish between three types of electronic transport regimes, depending on the parameters of the microscopic Hamiltonian, temperature and lattice size. 

 Hydrodynamic regime, discussed extensively in this work, occurs when momentum conserving collisions between electrons dominate over all other scattering processes. In this regime, the conductivity profiles follow the continuum description from kinetic theory, despite being measured in QMC on a discrete grid, the lattice.

 The hydrodynamic description is only valid if the electronic scattering length $l_{\mathrm{ee}}$ is significantly smaller than the width $W$ of the samples. In the reverse case, we enter ballistic regime where the electrons effectively do not interact within the sample's volume and the continuum description becomes inadequate. This is equivalent to a decrease of the scattering amplitude which we can modify by keeping only on-site Hubbard interactions in our microscopic Hamiltonian.
 
Simulating this system, yields a flat conductivity profile, shown in Fig.~\ref{fig:Hubbardandbeta10_a}, even for relatively large Hubbard interaction $U=3.44t$, which is already quite close to the critical coupling of the Mott transition on a hexagonal lattice. Thus we can conclude that it is possible to switch from hydrodynamic to ballistic transport by turning off the Coulomb tail of the electron-electron interaction at fixed temperature and system size.

  \begin{figure}[]
   \centering
   \subfigure[]     { \label{fig:Hubbardandbeta10_a}\includegraphics[width=0.49\textwidth]{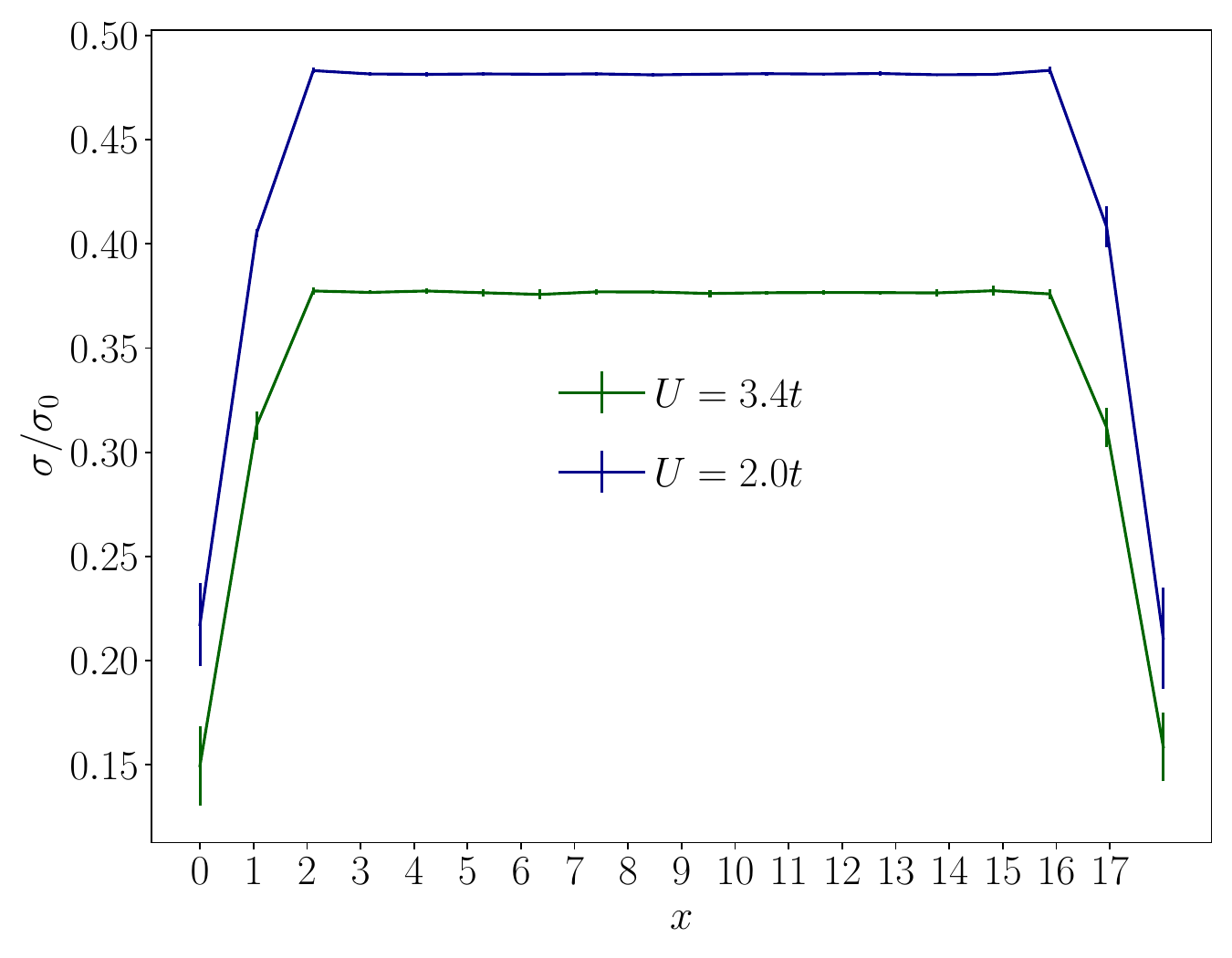}}
   \subfigure[]    { \label{fig:Hubbardandbeta10_b}\includegraphics[width=0.49\textwidth]{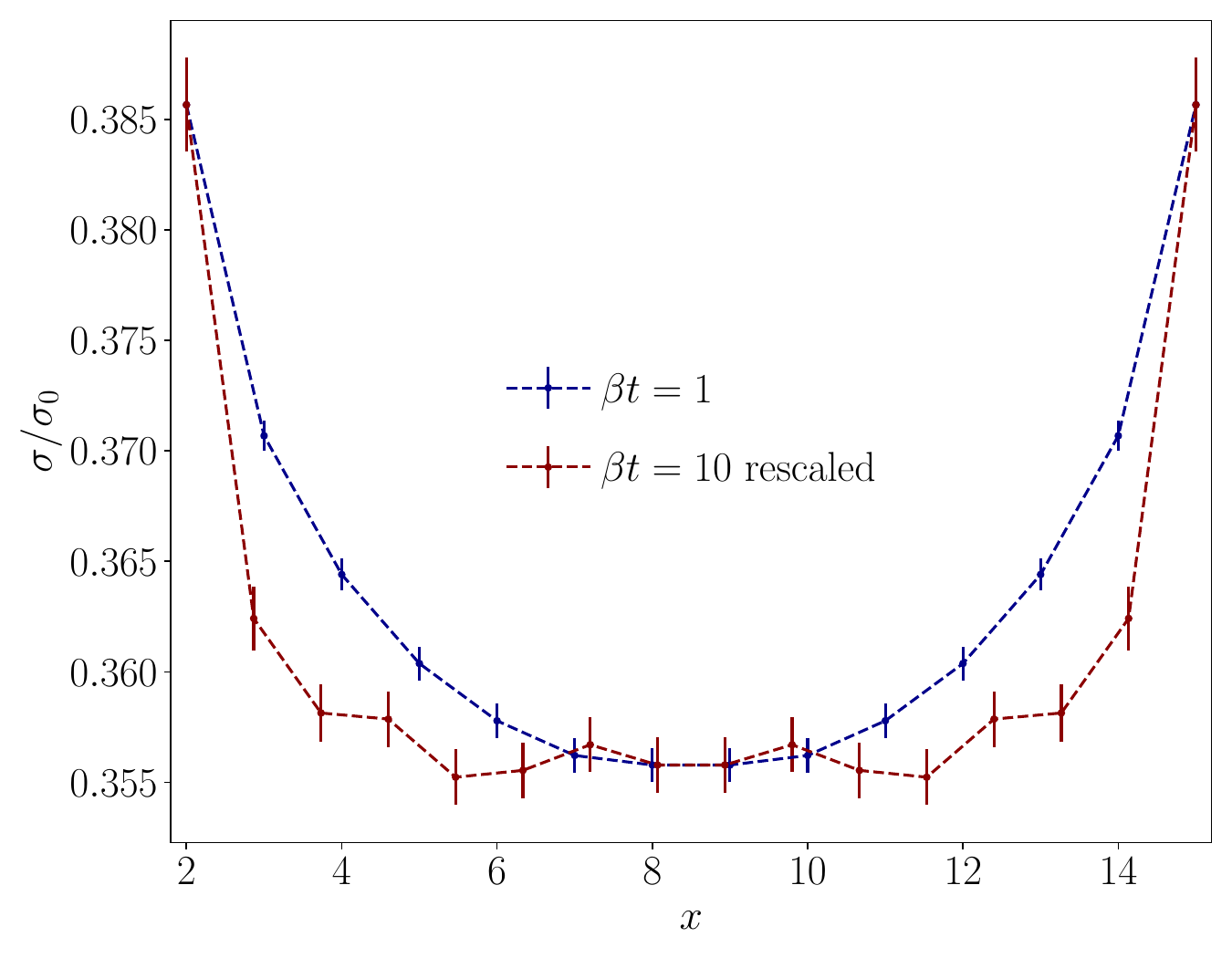}}
          \caption{(a) QMC conductivity profile for a system with only on-site Hubbard interactions $U=2t$ and $U =3.44t$. Temperature  is equal to hopping and lattice geometry is the same as the $T=t$ LRC run.  \\
    (b) Comparison between the QMC conductivity profiles for LRC runs on the lattice with 18 stripes at $T = 0.1 t$ and $T = t$. The profile of $T = 0.1t$ has been rescaled to match the edge and bulk conductivity of the $T = t$ profile.}
    \label{fig:Hubbardandbeta10}
 \end{figure}

 Choosing the correct temperature is also crucial for entering the hydrodynamic regime. The QMC scaling with system size puts a strong limit on possible system sizes. Consequently, in order to have enough charge carrier excitations contributing to the transport and therefore conductivity, the temperature needs to be high enough (in other words, we should enter the regime $\omega<T$).  Indeed, we can see in Fig.~\ref{fig:Hubbardandbeta10_b} that the QMC profiles, for a fixed lattice size but 10-times lower temperature, yields a flatter conductivity profile that does not follow the continuum description of kinetic theory. Thus the hydrodynamic transport can also be turned on and off by varying the temperature at a fixed interaction strength and system size, with lower temperatures leading to the suppression of hydrodynamic features of the electronic flow. 

Finally, the third regime, diffusive transport, does not conserve momentum and is dominant when $l_{\mathrm{ee}}$ is larger than the diffusive scattering length scale $l_{\mathrm{diff}}$ (and $l_{\mathrm{diff}} < W$).  Even though the simulation of this situation is out of scope of this paper, we can still 
 enter this regime by either enabling electron-phonon coupling in the microscopic Hamiltonian, or inserting adatoms over the whole lattice and not only the edges. With an increased density of adatoms, we expect the conductivity to flatten until reaching an ohmic type of profile.

\section{Coulomb interaction matrix}
For completeness, the effective Coulomb interaction matrix, taken from \cite{Wehling_2011} is plotted below (Fig. \ref{fig:CoulombPotential}) with the values of the onsite interaction $V_{00}$ and interactions in three next coordination radii written explicitly in the plot.

\begin{figure}[H]
        \centering
        \includegraphics[width=0.4\linewidth]{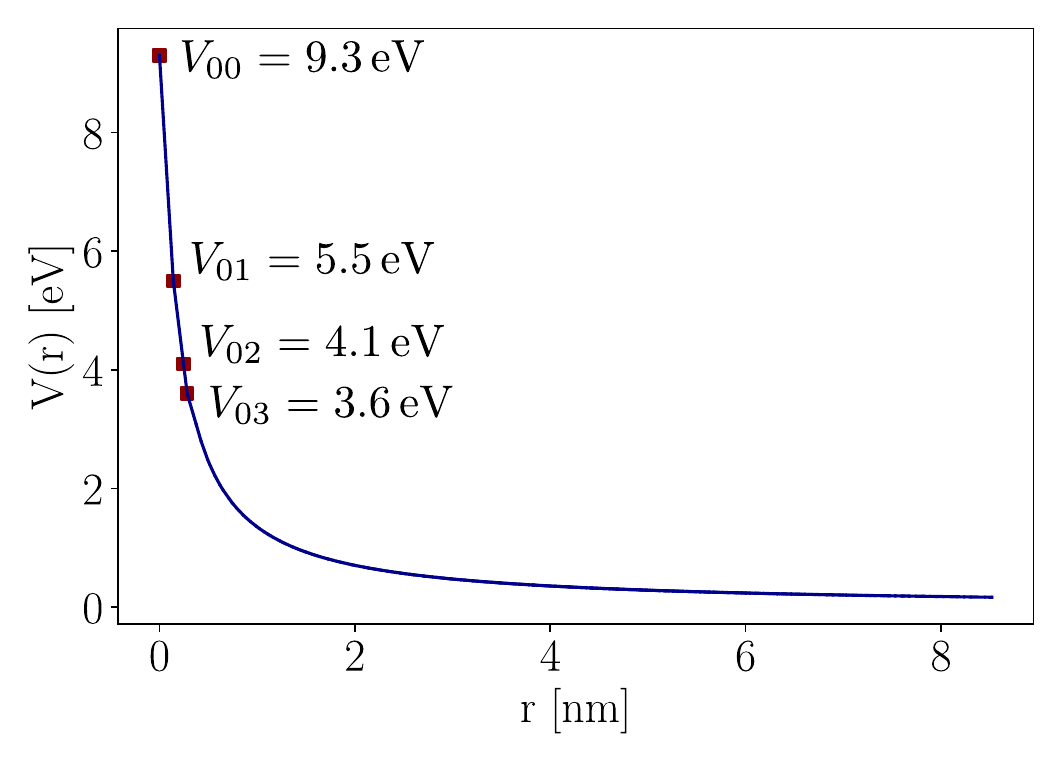}
    \caption{The effective Coulomb interaction in graphene taken from \cite{Wehling_2011}  with on-site potential $V_{00}$, nearest-neighbour potential $V_{01}$, next-nearest-neighbour potential $V_{02}$ and third-nearest-neighbour potential $V_{03}$. }
    \label{fig:CoulombPotential}
\end{figure}

\end{widetext}

\bibliography{general_bib, new_bib,fassaad.bib}

\begin{thebibliography}{50}%
\makeatletter
\providecommand \@ifxundefined [1]{%
 \@ifx{#1\undefined}
}%
\providecommand \@ifnum [1]{%
 \ifnum #1\expandafter \@firstoftwo
 \else \expandafter \@secondoftwo
 \fi
}%
\providecommand \@ifx [1]{%
 \ifx #1\expandafter \@firstoftwo
 \else \expandafter \@secondoftwo
 \fi
}%
\providecommand \natexlab [1]{#1}%
\providecommand \enquote  [1]{``#1''}%
\providecommand \bibnamefont  [1]{#1}%
\providecommand \bibfnamefont [1]{#1}%
\providecommand \citenamefont [1]{#1}%
\providecommand \href@noop [0]{\@secondoftwo}%
\providecommand \href [0]{\begingroup \@sanitize@url \@href}%
\providecommand \@href[1]{\@@startlink{#1}\@@href}%
\providecommand \@@href[1]{\endgroup#1\@@endlink}%
\providecommand \@sanitize@url [0]{\catcode `\\12\catcode `\$12\catcode `\&12\catcode `\#12\catcode `\^12\catcode `\_12\catcode `\%12\relax}%
\providecommand \@@startlink[1]{}%
\providecommand \@@endlink[0]{}%
\providecommand \url  [0]{\begingroup\@sanitize@url \@url }%
\providecommand \@url [1]{\endgroup\@href {#1}{\urlprefix }}%
\providecommand \urlprefix  [0]{URL }%
\providecommand \Eprint [0]{\href }%
\providecommand \doibase [0]{http://dx.doi.org/}%
\providecommand \selectlanguage [0]{\@gobble}%
\providecommand \bibinfo  [0]{\@secondoftwo}%
\providecommand \bibfield  [0]{\@secondoftwo}%
\providecommand \translation [1]{[#1]}%
\providecommand \BibitemOpen [0]{}%
\providecommand \bibitemStop [0]{}%
\providecommand \bibitemNoStop [0]{.\EOS\space}%
\providecommand \EOS [0]{\spacefactor3000\relax}%
\providecommand \BibitemShut  [1]{\csname bibitem#1\endcsname}%
\let\auto@bib@innerbib\@empty
\bibitem [{\citenamefont {Gurzhi}(1968)}]{Gurzhi68}%
  \BibitemOpen
  \bibfield  {author} {\bibinfo {author} {\bibfnamefont {R.~N.}\ \bibnamefont {Gurzhi}},\ }\href {\doibase 10.1070/PU1968v011n02ABEH003815} {\bibfield  {journal} {\bibinfo  {journal} {Soviet Physics Uspekhi}\ }\textbf {\bibinfo {volume} {11}},\ \bibinfo {pages} {255} (\bibinfo {year} {1968})}\BibitemShut {NoStop}%
\bibitem [{\citenamefont {de~Jong}\ and\ \citenamefont {Molenkamp}(1995)}]{deJong95}%
  \BibitemOpen
  \bibfield  {author} {\bibinfo {author} {\bibfnamefont {M.~J.~M.}\ \bibnamefont {de~Jong}}\ and\ \bibinfo {author} {\bibfnamefont {L.~W.}\ \bibnamefont {Molenkamp}},\ }\href {\doibase 10.1103/PhysRevB.51.13389} {\bibfield  {journal} {\bibinfo  {journal} {Phys. Rev. B}\ }\textbf {\bibinfo {volume} {51}},\ \bibinfo {pages} {13389} (\bibinfo {year} {1995})}\BibitemShut {NoStop}%
\bibitem [{\citenamefont {Damle}\ and\ \citenamefont {Sachdev}(1997)}]{PhysRevB.56.8714}%
  \BibitemOpen
  \bibfield  {author} {\bibinfo {author} {\bibfnamefont {K.}~\bibnamefont {Damle}}\ and\ \bibinfo {author} {\bibfnamefont {S.}~\bibnamefont {Sachdev}},\ }\href {\doibase 10.1103/PhysRevB.56.8714} {\bibfield  {journal} {\bibinfo  {journal} {Phys. Rev. B}\ }\textbf {\bibinfo {volume} {56}},\ \bibinfo {pages} {8714} (\bibinfo {year} {1997})}\BibitemShut {NoStop}%
\bibitem [{\citenamefont {M\"uller}\ \emph {et~al.}(2009{\natexlab{a}})\citenamefont {M\"uller}, \citenamefont {Schmalian},\ and\ \citenamefont {Fritz}}]{Fritz2009}%
  \BibitemOpen
  \bibfield  {author} {\bibinfo {author} {\bibfnamefont {M.}~\bibnamefont {M\"uller}}, \bibinfo {author} {\bibfnamefont {J.}~\bibnamefont {Schmalian}}, \ and\ \bibinfo {author} {\bibfnamefont {L.}~\bibnamefont {Fritz}},\ }\href {\doibase 10.1103/PhysRevLett.103.025301} {\bibfield  {journal} {\bibinfo  {journal} {Phys. Rev. Lett.}\ }\textbf {\bibinfo {volume} {103}},\ \bibinfo {pages} {025301} (\bibinfo {year} {2009}{\natexlab{a}})}\BibitemShut {NoStop}%
\bibitem [{\citenamefont {Lucas}\ and\ \citenamefont {Fong}(2018{\natexlab{a}})}]{Lucas_2018}%
  \BibitemOpen
  \bibfield  {author} {\bibinfo {author} {\bibfnamefont {A.}~\bibnamefont {Lucas}}\ and\ \bibinfo {author} {\bibfnamefont {K.~C.}\ \bibnamefont {Fong}},\ }\href {\doibase 10.1088/1361-648X/aaa274} {\bibfield  {journal} {\bibinfo  {journal} {Journal of Physics: Condensed Matter}\ }\textbf {\bibinfo {volume} {30}},\ \bibinfo {pages} {053001} (\bibinfo {year} {2018}{\natexlab{a}})}\BibitemShut {NoStop}%
\bibitem [{\citenamefont {Ho}\ \emph {et~al.}(2018)\citenamefont {Ho}, \citenamefont {Yudhistira}, \citenamefont {Chakraborty},\ and\ \citenamefont {Adam}}]{Ho18}%
  \BibitemOpen
  \bibfield  {author} {\bibinfo {author} {\bibfnamefont {D.~Y.~H.}\ \bibnamefont {Ho}}, \bibinfo {author} {\bibfnamefont {I.}~\bibnamefont {Yudhistira}}, \bibinfo {author} {\bibfnamefont {N.}~\bibnamefont {Chakraborty}}, \ and\ \bibinfo {author} {\bibfnamefont {S.}~\bibnamefont {Adam}},\ }\href {\doibase 10.1103/PhysRevB.97.121404} {\bibfield  {journal} {\bibinfo  {journal} {Phys. Rev. B}\ }\textbf {\bibinfo {volume} {97}},\ \bibinfo {pages} {121404} (\bibinfo {year} {2018})}\BibitemShut {NoStop}%
\bibitem [{\citenamefont {Levitov}\ and\ \citenamefont {Falkovich}(2016)}]{Levitov2016}%
  \BibitemOpen
  \bibfield  {author} {\bibinfo {author} {\bibfnamefont {L.}~\bibnamefont {Levitov}}\ and\ \bibinfo {author} {\bibfnamefont {G.}~\bibnamefont {Falkovich}},\ }\href {\doibase 10.1038/nphys3667} {\bibfield  {journal} {\bibinfo  {journal} {Nature Physics}\ }\textbf {\bibinfo {volume} {12}},\ \bibinfo {pages} {672} (\bibinfo {year} {2016})}\BibitemShut {NoStop}%
\bibitem [{\citenamefont {Karsch}\ \emph {et~al.}(2008)\citenamefont {Karsch}, \citenamefont {Kharzeev},\ and\ \citenamefont {Tuchin}}]{KARSCH2008217}%
  \BibitemOpen
  \bibfield  {author} {\bibinfo {author} {\bibfnamefont {F.}~\bibnamefont {Karsch}}, \bibinfo {author} {\bibfnamefont {D.}~\bibnamefont {Kharzeev}}, \ and\ \bibinfo {author} {\bibfnamefont {K.}~\bibnamefont {Tuchin}},\ }\href {\doibase https://doi.org/10.1016/j.physletb.2008.01.080} {\bibfield  {journal} {\bibinfo  {journal} {Physics Letters B}\ }\textbf {\bibinfo {volume} {663}},\ \bibinfo {pages} {217} (\bibinfo {year} {2008})}\BibitemShut {NoStop}%
\bibitem [{\citenamefont {Kovtun}\ \emph {et~al.}(2005)\citenamefont {Kovtun}, \citenamefont {Son},\ and\ \citenamefont {Starinets}}]{PhysRevLett.94.111601}%
  \BibitemOpen
  \bibfield  {author} {\bibinfo {author} {\bibfnamefont {P.~K.}\ \bibnamefont {Kovtun}}, \bibinfo {author} {\bibfnamefont {D.~T.}\ \bibnamefont {Son}}, \ and\ \bibinfo {author} {\bibfnamefont {A.~O.}\ \bibnamefont {Starinets}},\ }\href {\doibase 10.1103/PhysRevLett.94.111601} {\bibfield  {journal} {\bibinfo  {journal} {Phys. Rev. Lett.}\ }\textbf {\bibinfo {volume} {94}},\ \bibinfo {pages} {111601} (\bibinfo {year} {2005})}\BibitemShut {NoStop}%
\bibitem [{\citenamefont {Bandurin}\ \emph {et~al.}(2018)\citenamefont {Bandurin}, \citenamefont {Shytov}, \citenamefont {Levitov}, \citenamefont {Kumar}, \citenamefont {Berdyugin}, \citenamefont {Ben~Shalom}, \citenamefont {Grigorieva}, \citenamefont {Geim},\ and\ \citenamefont {Falkovich}}]{Bandurin2018}%
  \BibitemOpen
  \bibfield  {author} {\bibinfo {author} {\bibfnamefont {D.~A.}\ \bibnamefont {Bandurin}}, \bibinfo {author} {\bibfnamefont {A.~V.}\ \bibnamefont {Shytov}}, \bibinfo {author} {\bibfnamefont {L.~S.}\ \bibnamefont {Levitov}}, \bibinfo {author} {\bibfnamefont {R.~K.}\ \bibnamefont {Kumar}}, \bibinfo {author} {\bibfnamefont {A.~I.}\ \bibnamefont {Berdyugin}}, \bibinfo {author} {\bibfnamefont {M.}~\bibnamefont {Ben~Shalom}}, \bibinfo {author} {\bibfnamefont {I.~V.}\ \bibnamefont {Grigorieva}}, \bibinfo {author} {\bibfnamefont {A.~K.}\ \bibnamefont {Geim}}, \ and\ \bibinfo {author} {\bibfnamefont {G.}~\bibnamefont {Falkovich}},\ }\href {\doibase 10.1038/s41467-018-07004-4} {\bibfield  {journal} {\bibinfo  {journal} {Nature Communications}\ }\textbf {\bibinfo {volume} {9}},\ \bibinfo {pages} {4533} (\bibinfo {year} {2018})}\BibitemShut {NoStop}%
\bibitem [{\citenamefont {Sulpizio}\ \emph {et~al.}(2019)\citenamefont {Sulpizio}, \citenamefont {Ella}, \citenamefont {Rozen}, \citenamefont {Birkbeck}, \citenamefont {Perello}, \citenamefont {Dutta}, \citenamefont {Ben-Shalom}, \citenamefont {Taniguchi}, \citenamefont {Watanabe}, \citenamefont {Holder}, \citenamefont {Queiroz}, \citenamefont {Principi}, \citenamefont {Stern}, \citenamefont {Scaffidi}, \citenamefont {Geim},\ and\ \citenamefont {Ilani}}]{Sulpizio2019}%
  \BibitemOpen
  \bibfield  {author} {\bibinfo {author} {\bibfnamefont {J.~A.}\ \bibnamefont {Sulpizio}}, \bibinfo {author} {\bibfnamefont {L.}~\bibnamefont {Ella}}, \bibinfo {author} {\bibfnamefont {A.}~\bibnamefont {Rozen}}, \bibinfo {author} {\bibfnamefont {J.}~\bibnamefont {Birkbeck}}, \bibinfo {author} {\bibfnamefont {D.~J.}\ \bibnamefont {Perello}}, \bibinfo {author} {\bibfnamefont {D.}~\bibnamefont {Dutta}}, \bibinfo {author} {\bibfnamefont {M.}~\bibnamefont {Ben-Shalom}}, \bibinfo {author} {\bibfnamefont {T.}~\bibnamefont {Taniguchi}}, \bibinfo {author} {\bibfnamefont {K.}~\bibnamefont {Watanabe}}, \bibinfo {author} {\bibfnamefont {T.}~\bibnamefont {Holder}}, \bibinfo {author} {\bibfnamefont {R.}~\bibnamefont {Queiroz}}, \bibinfo {author} {\bibfnamefont {A.}~\bibnamefont {Principi}}, \bibinfo {author} {\bibfnamefont {A.}~\bibnamefont {Stern}}, \bibinfo {author} {\bibfnamefont {T.}~\bibnamefont {Scaffidi}}, \bibinfo {author} {\bibfnamefont {A.~K.}\ \bibnamefont {Geim}}, \ and\ \bibinfo {author} {\bibfnamefont
  {S.}~\bibnamefont {Ilani}},\ }\href {\doibase 10.1038/s41586-019-1788-9} {\bibfield  {journal} {\bibinfo  {journal} {Nature}\ }\textbf {\bibinfo {volume} {576}},\ \bibinfo {pages} {75} (\bibinfo {year} {2019})}\BibitemShut {NoStop}%
\bibitem [{\citenamefont {Ku}\ \emph {et~al.}(2020)\citenamefont {Ku}, \citenamefont {Zhou}, \citenamefont {Li}, \citenamefont {Shin}, \citenamefont {Shi}, \citenamefont {Burch}, \citenamefont {Anderson}, \citenamefont {Pierce}, \citenamefont {Xie}, \citenamefont {Hamo}, \citenamefont {Vool}, \citenamefont {Zhang}, \citenamefont {Casola}, \citenamefont {Taniguchi}, \citenamefont {Watanabe}, \citenamefont {Fogler}, \citenamefont {Kim}, \citenamefont {Yacoby},\ and\ \citenamefont {Walsworth}}]{Ku2020}%
  \BibitemOpen
  \bibfield  {author} {\bibinfo {author} {\bibfnamefont {M.~J.~H.}\ \bibnamefont {Ku}}, \bibinfo {author} {\bibfnamefont {T.~X.}\ \bibnamefont {Zhou}}, \bibinfo {author} {\bibfnamefont {Q.}~\bibnamefont {Li}}, \bibinfo {author} {\bibfnamefont {Y.~J.}\ \bibnamefont {Shin}}, \bibinfo {author} {\bibfnamefont {J.~K.}\ \bibnamefont {Shi}}, \bibinfo {author} {\bibfnamefont {C.}~\bibnamefont {Burch}}, \bibinfo {author} {\bibfnamefont {L.~E.}\ \bibnamefont {Anderson}}, \bibinfo {author} {\bibfnamefont {A.~T.}\ \bibnamefont {Pierce}}, \bibinfo {author} {\bibfnamefont {Y.}~\bibnamefont {Xie}}, \bibinfo {author} {\bibfnamefont {A.}~\bibnamefont {Hamo}}, \bibinfo {author} {\bibfnamefont {U.}~\bibnamefont {Vool}}, \bibinfo {author} {\bibfnamefont {H.}~\bibnamefont {Zhang}}, \bibinfo {author} {\bibfnamefont {F.}~\bibnamefont {Casola}}, \bibinfo {author} {\bibfnamefont {T.}~\bibnamefont {Taniguchi}}, \bibinfo {author} {\bibfnamefont {K.}~\bibnamefont {Watanabe}}, \bibinfo {author} {\bibfnamefont {M.~M.}\ \bibnamefont
  {Fogler}}, \bibinfo {author} {\bibfnamefont {P.}~\bibnamefont {Kim}}, \bibinfo {author} {\bibfnamefont {A.}~\bibnamefont {Yacoby}}, \ and\ \bibinfo {author} {\bibfnamefont {R.~L.}\ \bibnamefont {Walsworth}},\ }\href {\doibase 10.1038/s41586-020-2507-2} {\bibfield  {journal} {\bibinfo  {journal} {Nature}\ }\textbf {\bibinfo {volume} {583}},\ \bibinfo {pages} {537} (\bibinfo {year} {2020})}\BibitemShut {NoStop}%
\bibitem [{\citenamefont {Jenkins}\ \emph {et~al.}(2022)\citenamefont {Jenkins}, \citenamefont {Baumann}, \citenamefont {Zhou}, \citenamefont {Meynell}, \citenamefont {Daipeng}, \citenamefont {Watanabe}, \citenamefont {Taniguchi}, \citenamefont {Lucas}, \citenamefont {Young},\ and\ \citenamefont {Bleszynski~Jayich}}]{PhysRevLett.129.087701}%
  \BibitemOpen
  \bibfield  {author} {\bibinfo {author} {\bibfnamefont {A.}~\bibnamefont {Jenkins}}, \bibinfo {author} {\bibfnamefont {S.}~\bibnamefont {Baumann}}, \bibinfo {author} {\bibfnamefont {H.}~\bibnamefont {Zhou}}, \bibinfo {author} {\bibfnamefont {S.~A.}\ \bibnamefont {Meynell}}, \bibinfo {author} {\bibfnamefont {Y.}~\bibnamefont {Daipeng}}, \bibinfo {author} {\bibfnamefont {K.}~\bibnamefont {Watanabe}}, \bibinfo {author} {\bibfnamefont {T.}~\bibnamefont {Taniguchi}}, \bibinfo {author} {\bibfnamefont {A.}~\bibnamefont {Lucas}}, \bibinfo {author} {\bibfnamefont {A.~F.}\ \bibnamefont {Young}}, \ and\ \bibinfo {author} {\bibfnamefont {A.~C.}\ \bibnamefont {Bleszynski~Jayich}},\ }\href {\doibase 10.1103/PhysRevLett.129.087701} {\bibfield  {journal} {\bibinfo  {journal} {Phys. Rev. Lett.}\ }\textbf {\bibinfo {volume} {129}},\ \bibinfo {pages} {087701} (\bibinfo {year} {2022})}\BibitemShut {NoStop}%
\bibitem [{\citenamefont {Narozhny}\ \emph {et~al.}(2021)\citenamefont {Narozhny}, \citenamefont {Gornyi},\ and\ \citenamefont {Titov}}]{Narozhny_2021}%
  \BibitemOpen
  \bibfield  {author} {\bibinfo {author} {\bibfnamefont {B.~N.}\ \bibnamefont {Narozhny}}, \bibinfo {author} {\bibfnamefont {I.~V.}\ \bibnamefont {Gornyi}}, \ and\ \bibinfo {author} {\bibfnamefont {M.}~\bibnamefont {Titov}},\ }\href {\doibase 10.1103/PhysRevB.104.075443} {\bibfield  {journal} {\bibinfo  {journal} {Phys. Rev. B}\ }\textbf {\bibinfo {volume} {104}},\ \bibinfo {pages} {075443} (\bibinfo {year} {2021})}\BibitemShut {NoStop}%
\bibitem [{Sup()}]{Supplement}%
  \BibitemOpen
  \href@noop {} {}\bibinfo {note} {See Supplemental Material at [URL], which includes Refs. [48-50] for additional information about the kinetic theory and QMC simulations.}\BibitemShut {Stop}%
\bibitem [{\citenamefont {Sorella}\ and\ \citenamefont {Tosatti}(1992)}]{Sorella_1992}%
  \BibitemOpen
  \bibfield  {author} {\bibinfo {author} {\bibfnamefont {S.}~\bibnamefont {Sorella}}\ and\ \bibinfo {author} {\bibfnamefont {E.}~\bibnamefont {Tosatti}},\ }\href {\doibase 10.1209/0295-5075/19/8/007} {\bibfield  {journal} {\bibinfo  {journal} {Europhysics Letters}\ }\textbf {\bibinfo {volume} {19}},\ \bibinfo {pages} {699} (\bibinfo {year} {1992})}\BibitemShut {NoStop}%
\bibitem [{\citenamefont {Troyer}\ and\ \citenamefont {Wiese}(2005)}]{Troyer05}%
  \BibitemOpen
  \bibfield  {author} {\bibinfo {author} {\bibfnamefont {M.}~\bibnamefont {Troyer}}\ and\ \bibinfo {author} {\bibfnamefont {U.-J.}\ \bibnamefont {Wiese}},\ }\href {\doibase 10.1103/PhysRevLett.94.170201} {\bibfield  {journal} {\bibinfo  {journal} {Phys. Rev. Lett.}\ }\textbf {\bibinfo {volume} {94}},\ \bibinfo {pages} {170201} (\bibinfo {year} {2005})}\BibitemShut {NoStop}%
\bibitem [{\citenamefont {Ulybyshev}\ \emph {et~al.}(2013)\citenamefont {Ulybyshev}, \citenamefont {Buividovich}, \citenamefont {Katsnelson},\ and\ \citenamefont {Polikarpov}}]{Ulybyshev13}%
  \BibitemOpen
  \bibfield  {author} {\bibinfo {author} {\bibfnamefont {M.~V.}\ \bibnamefont {Ulybyshev}}, \bibinfo {author} {\bibfnamefont {P.~V.}\ \bibnamefont {Buividovich}}, \bibinfo {author} {\bibfnamefont {M.~I.}\ \bibnamefont {Katsnelson}}, \ and\ \bibinfo {author} {\bibfnamefont {M.~I.}\ \bibnamefont {Polikarpov}},\ }\href {\doibase 10.1103/PhysRevLett.111.056801} {\bibfield  {journal} {\bibinfo  {journal} {Phys. Rev. Lett.}\ }\textbf {\bibinfo {volume} {111}},\ \bibinfo {pages} {056801} (\bibinfo {year} {2013})}\BibitemShut {NoStop}%
\bibitem [{\citenamefont {Hohenadler}\ \emph {et~al.}(2014)\citenamefont {Hohenadler}, \citenamefont {Parisen~Toldin}, \citenamefont {Herbut},\ and\ \citenamefont {Assaad}}]{Hohenadler14}%
  \BibitemOpen
  \bibfield  {author} {\bibinfo {author} {\bibfnamefont {M.}~\bibnamefont {Hohenadler}}, \bibinfo {author} {\bibfnamefont {F.}~\bibnamefont {Parisen~Toldin}}, \bibinfo {author} {\bibfnamefont {I.~F.}\ \bibnamefont {Herbut}}, \ and\ \bibinfo {author} {\bibfnamefont {F.~F.}\ \bibnamefont {Assaad}},\ }\href {\doibase 10.1103/PhysRevB.90.085146} {\bibfield  {journal} {\bibinfo  {journal} {Phys. Rev. B}\ }\textbf {\bibinfo {volume} {90}},\ \bibinfo {pages} {085146} (\bibinfo {year} {2014})}\BibitemShut {NoStop}%
\bibitem [{\citenamefont {Jim{\'e}nez}\ \emph {et~al.}(2021)\citenamefont {Jim{\'e}nez}, \citenamefont {Crone}, \citenamefont {Fogh}, \citenamefont {Zayed}, \citenamefont {Lortz}, \citenamefont {Pomjakushina}, \citenamefont {Conder}, \citenamefont {L{\"a}uchli}, \citenamefont {Weber}, \citenamefont {Wessel}, \citenamefont {Honecker}, \citenamefont {Normand}, \citenamefont {R{\"u}egg}, \citenamefont {Corboz}, \citenamefont {R{\o}nnow},\ and\ \citenamefont {Mila}}]{Jiménez2021}%
  \BibitemOpen
  \bibfield  {author} {\bibinfo {author} {\bibfnamefont {J.~L.}\ \bibnamefont {Jim{\'e}nez}}, \bibinfo {author} {\bibfnamefont {S.~P.~G.}\ \bibnamefont {Crone}}, \bibinfo {author} {\bibfnamefont {E.}~\bibnamefont {Fogh}}, \bibinfo {author} {\bibfnamefont {M.~E.}\ \bibnamefont {Zayed}}, \bibinfo {author} {\bibfnamefont {R.}~\bibnamefont {Lortz}}, \bibinfo {author} {\bibfnamefont {E.}~\bibnamefont {Pomjakushina}}, \bibinfo {author} {\bibfnamefont {K.}~\bibnamefont {Conder}}, \bibinfo {author} {\bibfnamefont {A.~M.}\ \bibnamefont {L{\"a}uchli}}, \bibinfo {author} {\bibfnamefont {L.}~\bibnamefont {Weber}}, \bibinfo {author} {\bibfnamefont {S.}~\bibnamefont {Wessel}}, \bibinfo {author} {\bibfnamefont {A.}~\bibnamefont {Honecker}}, \bibinfo {author} {\bibfnamefont {B.}~\bibnamefont {Normand}}, \bibinfo {author} {\bibfnamefont {C.}~\bibnamefont {R{\"u}egg}}, \bibinfo {author} {\bibfnamefont {P.}~\bibnamefont {Corboz}}, \bibinfo {author} {\bibfnamefont {H.~M.}\ \bibnamefont {R{\o}nnow}}, \ and\ \bibinfo {author}
  {\bibfnamefont {F.}~\bibnamefont {Mila}},\ }\href {\doibase 10.1038/s41586-021-03411-8} {\bibfield  {journal} {\bibinfo  {journal} {Nature}\ }\textbf {\bibinfo {volume} {592}},\ \bibinfo {pages} {370} (\bibinfo {year} {2021})}\BibitemShut {NoStop}%
\bibitem [{\citenamefont {Kovtun}(2012)}]{Kovtun_2012}%
  \BibitemOpen
  \bibfield  {author} {\bibinfo {author} {\bibfnamefont {P.}~\bibnamefont {Kovtun}},\ }\href {\doibase 10.1088/1751-8113/45/47/473001} {\bibfield  {journal} {\bibinfo  {journal} {Journal of Physics A: Mathematical and Theoretical}\ }\textbf {\bibinfo {volume} {45}},\ \bibinfo {pages} {473001} (\bibinfo {year} {2012})}\BibitemShut {NoStop}%
\bibitem [{\citenamefont {Lucas}\ and\ \citenamefont {Fong}(2018{\natexlab{b}})}]{Lucas2018}%
  \BibitemOpen
  \bibfield  {author} {\bibinfo {author} {\bibfnamefont {A.}~\bibnamefont {Lucas}}\ and\ \bibinfo {author} {\bibfnamefont {K.~C.}\ \bibnamefont {Fong}},\ }\href {\doibase 10.1088/1361-648X/aaa274} {\bibfield  {journal} {\bibinfo  {journal} {Journal of Physics: Condensed Matter}\ }\textbf {\bibinfo {volume} {30}},\ \bibinfo {pages} {053001} (\bibinfo {year} {2018}{\natexlab{b}})}\BibitemShut {NoStop}%
\bibitem [{\citenamefont {Narozhny}(2019)}]{Narozhny2019}%
  \BibitemOpen
  \bibfield  {author} {\bibinfo {author} {\bibfnamefont {B.~N.}\ \bibnamefont {Narozhny}},\ }\href {\doibase https://doi.org/10.1016/j.aop.2019.167979} {\bibfield  {journal} {\bibinfo  {journal} {Annals of Physics}\ }\textbf {\bibinfo {volume} {411}},\ \bibinfo {pages} {167979} (\bibinfo {year} {2019})}\BibitemShut {NoStop}%
\bibitem [{\citenamefont {Fritz}\ and\ \citenamefont {Scaffidi}(2024)}]{Fritz2024}%
  \BibitemOpen
  \bibfield  {author} {\bibinfo {author} {\bibfnamefont {L.}~\bibnamefont {Fritz}}\ and\ \bibinfo {author} {\bibfnamefont {T.}~\bibnamefont {Scaffidi}},\ }\href {\doibase https://doi.org/10.1146/annurev-conmatphys-040521-042014} {\bibfield  {journal} {\bibinfo  {journal} {Annual Review of Condensed Matter Physics}\ }\textbf {\bibinfo {volume} {15}},\ \bibinfo {pages} {17} (\bibinfo {year} {2024})}\BibitemShut {NoStop}%
\bibitem [{\citenamefont {Pongsangangan}\ \emph {et~al.}(2022)\citenamefont {Pongsangangan}, \citenamefont {Ludwig}, \citenamefont {Stoof},\ and\ \citenamefont {Fritz}}]{Pongsangangan2022}%
  \BibitemOpen
  \bibfield  {author} {\bibinfo {author} {\bibfnamefont {K.}~\bibnamefont {Pongsangangan}}, \bibinfo {author} {\bibfnamefont {T.}~\bibnamefont {Ludwig}}, \bibinfo {author} {\bibfnamefont {H.~T.~C.}\ \bibnamefont {Stoof}}, \ and\ \bibinfo {author} {\bibfnamefont {L.}~\bibnamefont {Fritz}},\ }\href {\doibase 10.1103/PhysRevB.106.205126} {\bibfield  {journal} {\bibinfo  {journal} {Phys. Rev. B}\ }\textbf {\bibinfo {volume} {106}},\ \bibinfo {pages} {205126} (\bibinfo {year} {2022})}\BibitemShut {NoStop}%
\bibitem [{\citenamefont {Fritz}\ \emph {et~al.}(2008)\citenamefont {Fritz}, \citenamefont {Schmalian}, \citenamefont {M\"uller},\ and\ \citenamefont {Sachdev}}]{Fritz_2008}%
  \BibitemOpen
  \bibfield  {author} {\bibinfo {author} {\bibfnamefont {L.}~\bibnamefont {Fritz}}, \bibinfo {author} {\bibfnamefont {J.}~\bibnamefont {Schmalian}}, \bibinfo {author} {\bibfnamefont {M.}~\bibnamefont {M\"uller}}, \ and\ \bibinfo {author} {\bibfnamefont {S.}~\bibnamefont {Sachdev}},\ }\href {\doibase 10.1103/PhysRevB.78.085416} {\bibfield  {journal} {\bibinfo  {journal} {Phys. Rev. B}\ }\textbf {\bibinfo {volume} {78}},\ \bibinfo {pages} {085416} (\bibinfo {year} {2008})}\BibitemShut {NoStop}%
\bibitem [{\citenamefont {Ziman}(2001)}]{Ziman2001}%
  \BibitemOpen
  \bibfield  {author} {\bibinfo {author} {\bibfnamefont {J.}~\bibnamefont {Ziman}},\ }\href {\doibase 10.1093/acprof:oso/9780198507796.001.0001} {\emph {\bibinfo {title} {Electrons and Phonons: The Theory of Transport Phenomena in Solids}}}\ (\bibinfo  {publisher} {Oxford University Press},\ \bibinfo {year} {2001})\BibitemShut {NoStop}%
\bibitem [{\citenamefont {Pongsangangan}\ \emph {et~al.}(2024)\citenamefont {Pongsangangan}, \citenamefont {Cosme}, \citenamefont {Di~Salvo},\ and\ \citenamefont {Fritz}}]{Kitinan2024}%
  \BibitemOpen
  \bibfield  {author} {\bibinfo {author} {\bibfnamefont {K.}~\bibnamefont {Pongsangangan}}, \bibinfo {author} {\bibfnamefont {P.}~\bibnamefont {Cosme}}, \bibinfo {author} {\bibfnamefont {E.}~\bibnamefont {Di~Salvo}}, \ and\ \bibinfo {author} {\bibfnamefont {L.}~\bibnamefont {Fritz}},\ }\href {\doibase 10.1103/PhysRevB.110.045443} {\bibfield  {journal} {\bibinfo  {journal} {Phys. Rev. B}\ }\textbf {\bibinfo {volume} {110}},\ \bibinfo {pages} {045443} (\bibinfo {year} {2024})}\BibitemShut {NoStop}%
\bibitem [{\citenamefont {Kiselev}\ and\ \citenamefont {Schmalian}(2019)}]{Kiselev2019}%
  \BibitemOpen
  \bibfield  {author} {\bibinfo {author} {\bibfnamefont {E.~I.}\ \bibnamefont {Kiselev}}\ and\ \bibinfo {author} {\bibfnamefont {J.}~\bibnamefont {Schmalian}},\ }\href {\doibase 10.1103/PhysRevB.99.035430} {\bibfield  {journal} {\bibinfo  {journal} {Phys. Rev. B}\ }\textbf {\bibinfo {volume} {99}},\ \bibinfo {pages} {035430} (\bibinfo {year} {2019})}\BibitemShut {NoStop}%
\bibitem [{\citenamefont {Wehling}\ \emph {et~al.}(2011)\citenamefont {Wehling}, \citenamefont {\ifmmode \mbox{\c{S}}\else \c{S}\fi{}a\ifmmode \mbox{\c{s}}\else \c{s}\fi{}\ifmmode \imath \else \i \fi{}o\ifmmode~\breve{g}\else \u{g}\fi{}lu}, \citenamefont {Friedrich}, \citenamefont {Lichtenstein}, \citenamefont {Katsnelson},\ and\ \citenamefont {Bl\"ugel}}]{Wehling_2011}%
  \BibitemOpen
  \bibfield  {author} {\bibinfo {author} {\bibfnamefont {T.~O.}\ \bibnamefont {Wehling}}, \bibinfo {author} {\bibfnamefont {E.}~\bibnamefont {\ifmmode \mbox{\c{S}}\else \c{S}\fi{}a\ifmmode \mbox{\c{s}}\else \c{s}\fi{}\ifmmode \imath \else \i \fi{}o\ifmmode~\breve{g}\else \u{g}\fi{}lu}}, \bibinfo {author} {\bibfnamefont {C.}~\bibnamefont {Friedrich}}, \bibinfo {author} {\bibfnamefont {A.~I.}\ \bibnamefont {Lichtenstein}}, \bibinfo {author} {\bibfnamefont {M.~I.}\ \bibnamefont {Katsnelson}}, \ and\ \bibinfo {author} {\bibfnamefont {S.}~\bibnamefont {Bl\"ugel}},\ }\href {\doibase 10.1103/PhysRevLett.106.236805} {\bibfield  {journal} {\bibinfo  {journal} {Phys. Rev. Lett.}\ }\textbf {\bibinfo {volume} {106}},\ \bibinfo {pages} {236805} (\bibinfo {year} {2011})}\BibitemShut {NoStop}%
\bibitem [{\citenamefont {Ulybyshev}\ \emph {et~al.}(2021)\citenamefont {Ulybyshev}, \citenamefont {Zafeiropoulos}, \citenamefont {Winterowd},\ and\ \citenamefont {Assaad}}]{Ulybyshev:2021xao}%
  \BibitemOpen
  \bibfield  {author} {\bibinfo {author} {\bibfnamefont {M.}~\bibnamefont {Ulybyshev}}, \bibinfo {author} {\bibfnamefont {S.}~\bibnamefont {Zafeiropoulos}}, \bibinfo {author} {\bibfnamefont {C.}~\bibnamefont {Winterowd}}, \ and\ \bibinfo {author} {\bibfnamefont {F.}~\bibnamefont {Assaad}},\ }\href@noop {} {\  (\bibinfo {year} {2021})},\ \Eprint {http://arxiv.org/abs/2104.09655} {arXiv:2104.09655 [cond-mat.str-el]} \BibitemShut {NoStop}%
\bibitem [{\citenamefont {Ulybyshev}\ and\ \citenamefont {Katsnelson}(2015)}]{PhysRevLett.114.246801}%
  \BibitemOpen
  \bibfield  {author} {\bibinfo {author} {\bibfnamefont {M.~V.}\ \bibnamefont {Ulybyshev}}\ and\ \bibinfo {author} {\bibfnamefont {M.~I.}\ \bibnamefont {Katsnelson}},\ }\href {\doibase 10.1103/PhysRevLett.114.246801} {\bibfield  {journal} {\bibinfo  {journal} {Phys. Rev. Lett.}\ }\textbf {\bibinfo {volume} {114}},\ \bibinfo {pages} {246801} (\bibinfo {year} {2015})}\BibitemShut {NoStop}%
\bibitem [{\citenamefont {Blankenbecler}\ \emph {et~al.}(1981)\citenamefont {Blankenbecler}, \citenamefont {Scalapino},\ and\ \citenamefont {Sugar}}]{Blankenbecler81}%
  \BibitemOpen
  \bibfield  {author} {\bibinfo {author} {\bibfnamefont {R.}~\bibnamefont {Blankenbecler}}, \bibinfo {author} {\bibfnamefont {D.~J.}\ \bibnamefont {Scalapino}}, \ and\ \bibinfo {author} {\bibfnamefont {R.~L.}\ \bibnamefont {Sugar}},\ }\href {\doibase 10.1103/PhysRevD.24.2278} {\bibfield  {journal} {\bibinfo  {journal} {Phys. Rev. D}\ }\textbf {\bibinfo {volume} {24}},\ \bibinfo {pages} {2278} (\bibinfo {year} {1981})}\BibitemShut {NoStop}%
\bibitem [{\citenamefont {White}\ \emph {et~al.}(1989)\citenamefont {White}, \citenamefont {Scalapino}, \citenamefont {Sugar}, \citenamefont {Loh}, \citenamefont {Gubernatis},\ and\ \citenamefont {Scalettar}}]{White89}%
  \BibitemOpen
  \bibfield  {author} {\bibinfo {author} {\bibfnamefont {S.}~\bibnamefont {White}}, \bibinfo {author} {\bibfnamefont {D.}~\bibnamefont {Scalapino}}, \bibinfo {author} {\bibfnamefont {R.}~\bibnamefont {Sugar}}, \bibinfo {author} {\bibfnamefont {E.}~\bibnamefont {Loh}}, \bibinfo {author} {\bibfnamefont {J.}~\bibnamefont {Gubernatis}}, \ and\ \bibinfo {author} {\bibfnamefont {R.}~\bibnamefont {Scalettar}},\ }\href {\doibase 10.1103/PhysRevB.40.506} {\bibfield  {journal} {\bibinfo  {journal} {Phys. Rev. B}\ }\textbf {\bibinfo {volume} {40}},\ \bibinfo {pages} {506} (\bibinfo {year} {1989})}\BibitemShut {NoStop}%
\bibitem [{\citenamefont {Assaad}\ and\ \citenamefont {Evertz}(2008)}]{Assaad08_rev}%
  \BibitemOpen
  \bibfield  {author} {\bibinfo {author} {\bibfnamefont {F.}~\bibnamefont {Assaad}}\ and\ \bibinfo {author} {\bibfnamefont {H.}~\bibnamefont {Evertz}},\ }in\ \href {\doibase 10.1007/978-3-540-74686-7_10} {\emph {\bibinfo {booktitle} {Computational Many-Particle Physics}}},\ \bibinfo {series} {Lecture Notes in Physics}, Vol.\ \bibinfo {volume} {739},\ \bibinfo {editor} {edited by\ \bibinfo {editor} {\bibfnamefont {H.}~\bibnamefont {Fehske}}, \bibinfo {editor} {\bibfnamefont {R.}~\bibnamefont {Schneider}}, \ and\ \bibinfo {editor} {\bibfnamefont {A.}~\bibnamefont {Wei{\ss}e}}}\ (\bibinfo  {publisher} {Springer},\ \bibinfo {address} {Berlin Heidelberg},\ \bibinfo {year} {2008})\ pp.\ \bibinfo {pages} {277--356}\BibitemShut {NoStop}%
\bibitem [{\citenamefont {Assaad}\ \emph {et~al.}(2022)\citenamefont {Assaad}, \citenamefont {Bercx}, \citenamefont {Goth}, \citenamefont {G{\"o}tz}, \citenamefont {Hofmann}, \citenamefont {Huffman}, \citenamefont {Liu}, \citenamefont {Toldin}, \citenamefont {Portela},\ and\ \citenamefont {Schwab}}]{ALF_v2}%
  \BibitemOpen
  \bibfield  {author} {\bibinfo {author} {\bibfnamefont {F.~F.}\ \bibnamefont {Assaad}}, \bibinfo {author} {\bibfnamefont {M.}~\bibnamefont {Bercx}}, \bibinfo {author} {\bibfnamefont {F.}~\bibnamefont {Goth}}, \bibinfo {author} {\bibfnamefont {A.}~\bibnamefont {G{\"o}tz}}, \bibinfo {author} {\bibfnamefont {J.~S.}\ \bibnamefont {Hofmann}}, \bibinfo {author} {\bibfnamefont {E.}~\bibnamefont {Huffman}}, \bibinfo {author} {\bibfnamefont {Z.}~\bibnamefont {Liu}}, \bibinfo {author} {\bibfnamefont {F.~P.}\ \bibnamefont {Toldin}}, \bibinfo {author} {\bibfnamefont {J.~S.~E.}\ \bibnamefont {Portela}}, \ and\ \bibinfo {author} {\bibfnamefont {J.}~\bibnamefont {Schwab}},\ }\href {\doibase 10.21468/SciPostPhysCodeb.1} {\bibfield  {journal} {\bibinfo  {journal} {SciPost Phys. Codebases}\ ,\ \bibinfo {pages} {1}} (\bibinfo {year} {2022})}\BibitemShut {NoStop}%
\bibitem [{\citenamefont {{Beach}}(2004)}]{Beach04a}%
  \BibitemOpen
  \bibfield  {author} {\bibinfo {author} {\bibfnamefont {K.~S.~D.}\ \bibnamefont {{Beach}}},\ }\href {\doibase 10.48550/arXiv.cond-mat/0403055} {\bibfield  {journal} {\bibinfo  {journal} {arXiv e-prints}\ ,\ \bibinfo {eid} {cond-mat/0403055}} (\bibinfo {year} {2004})},\ \Eprint {http://arxiv.org/abs/cond-mat/0403055} {arXiv:cond-mat/0403055 [cond-mat.str-el]} \BibitemShut {NoStop}%
\bibitem [{\citenamefont {Sandvik}(1998)}]{Sandvik98}%
  \BibitemOpen
  \bibfield  {author} {\bibinfo {author} {\bibfnamefont {A.}~\bibnamefont {Sandvik}},\ }\href {\doibase 10.1103/PhysRevB.57.10287} {\bibfield  {journal} {\bibinfo  {journal} {Phys. Rev. B}\ }\textbf {\bibinfo {volume} {57}},\ \bibinfo {pages} {10287} (\bibinfo {year} {1998})}\BibitemShut {NoStop}%
\bibitem [{\citenamefont {Shao}\ and\ \citenamefont {Sandvik}(2023)}]{SHAO20231}%
  \BibitemOpen
  \bibfield  {author} {\bibinfo {author} {\bibfnamefont {H.}~\bibnamefont {Shao}}\ and\ \bibinfo {author} {\bibfnamefont {A.~W.}\ \bibnamefont {Sandvik}},\ }\href {\doibase https://doi.org/10.1016/j.physrep.2022.11.002} {\bibfield  {journal} {\bibinfo  {journal} {Physics Reports}\ }\textbf {\bibinfo {volume} {1003}},\ \bibinfo {pages} {1} (\bibinfo {year} {2023})},\ \bibinfo {note} {progress on stochastic analytic continuation of quantum Monte Carlo data}\BibitemShut {NoStop}%
\bibitem [{\citenamefont {Mahan}(1990)}]{Mahan90}%
  \BibitemOpen
  \bibfield  {author} {\bibinfo {author} {\bibfnamefont {G.~D.}\ \bibnamefont {Mahan}},\ }\href@noop {} {\emph {\bibinfo {title} {Many-Particle Physics}}},\ \bibinfo {edition} {2nd}\ ed.\ (\bibinfo  {publisher} {Plenum Press},\ \bibinfo {address} {New York},\ \bibinfo {year} {1990})\BibitemShut {NoStop}%
\bibitem [{\citenamefont {Kubo}(1957)}]{Kubo57}%
  \BibitemOpen
  \bibfield  {author} {\bibinfo {author} {\bibfnamefont {R.}~\bibnamefont {Kubo}},\ }\href {\doibase 10.1143/JPSJ.12.570} {\bibfield  {journal} {\bibinfo  {journal} {Journal of the Physical Society of Japan}\ }\textbf {\bibinfo {volume} {12}},\ \bibinfo {pages} {570} (\bibinfo {year} {1957})},\ \Eprint {http://arxiv.org/abs/https://doi.org/10.1143/JPSJ.12.570} {https://doi.org/10.1143/JPSJ.12.570} \BibitemShut {NoStop}%
\bibitem [{\citenamefont {Jarrell}\ and\ \citenamefont {Gubernatis}(1996)}]{Jarrell96}%
  \BibitemOpen
  \bibfield  {author} {\bibinfo {author} {\bibfnamefont {M.}~\bibnamefont {Jarrell}}\ and\ \bibinfo {author} {\bibfnamefont {J.}~\bibnamefont {Gubernatis}},\ }\href@noop {} {\bibfield  {journal} {\bibinfo  {journal} {Physics Reports}\ }\textbf {\bibinfo {volume} {269}},\ \bibinfo {pages} {133} (\bibinfo {year} {1996})}\BibitemShut {NoStop}%
\bibitem [{\citenamefont {Stauber}\ \emph {et~al.}(2017)\citenamefont {Stauber}, \citenamefont {Parida}, \citenamefont {Trushin}, \citenamefont {Ulybyshev}, \citenamefont {Boyda},\ and\ \citenamefont {Schliemann}}]{PhysRevLett.118.266801}%
  \BibitemOpen
  \bibfield  {author} {\bibinfo {author} {\bibfnamefont {T.}~\bibnamefont {Stauber}}, \bibinfo {author} {\bibfnamefont {P.}~\bibnamefont {Parida}}, \bibinfo {author} {\bibfnamefont {M.}~\bibnamefont {Trushin}}, \bibinfo {author} {\bibfnamefont {M.~V.}\ \bibnamefont {Ulybyshev}}, \bibinfo {author} {\bibfnamefont {D.~L.}\ \bibnamefont {Boyda}}, \ and\ \bibinfo {author} {\bibfnamefont {J.}~\bibnamefont {Schliemann}},\ }\href {\doibase 10.1103/PhysRevLett.118.266801} {\bibfield  {journal} {\bibinfo  {journal} {Phys. Rev. Lett.}\ }\textbf {\bibinfo {volume} {118}},\ \bibinfo {pages} {266801} (\bibinfo {year} {2017})}\BibitemShut {NoStop}%
\bibitem [{\citenamefont {Trivedi}\ \emph {et~al.}(1996)\citenamefont {Trivedi}, \citenamefont {Scalettar},\ and\ \citenamefont {Randeria}}]{Trivedi1996}%
  \BibitemOpen
  \bibfield  {author} {\bibinfo {author} {\bibfnamefont {N.}~\bibnamefont {Trivedi}}, \bibinfo {author} {\bibfnamefont {R.~T.}\ \bibnamefont {Scalettar}}, \ and\ \bibinfo {author} {\bibfnamefont {M.}~\bibnamefont {Randeria}},\ }\href {\doibase 10.1103/PhysRevB.54.R3756} {\bibfield  {journal} {\bibinfo  {journal} {Phys. Rev. B}\ }\textbf {\bibinfo {volume} {54}},\ \bibinfo {pages} {R3756} (\bibinfo {year} {1996})}\BibitemShut {NoStop}%
\bibitem [{\citenamefont {Boyda}\ \emph {et~al.}(2016)\citenamefont {Boyda}, \citenamefont {Braguta}, \citenamefont {Katsnelson},\ and\ \citenamefont {Ulybyshev}}]{PhysRevB.94.085421}%
  \BibitemOpen
  \bibfield  {author} {\bibinfo {author} {\bibfnamefont {D.~L.}\ \bibnamefont {Boyda}}, \bibinfo {author} {\bibfnamefont {V.~V.}\ \bibnamefont {Braguta}}, \bibinfo {author} {\bibfnamefont {M.~I.}\ \bibnamefont {Katsnelson}}, \ and\ \bibinfo {author} {\bibfnamefont {M.~V.}\ \bibnamefont {Ulybyshev}},\ }\href {\doibase 10.1103/PhysRevB.94.085421} {\bibfield  {journal} {\bibinfo  {journal} {Phys. Rev. B}\ }\textbf {\bibinfo {volume} {94}},\ \bibinfo {pages} {085421} (\bibinfo {year} {2016})}\BibitemShut {NoStop}%
\bibitem [{\citenamefont {Reingruber}(2024)}]{Data}%
  \BibitemOpen
  \bibfield  {author} {\bibinfo {author} {\bibfnamefont {A.}~\bibnamefont {Reingruber}},\ }\href@noop {} {\enquote {\bibinfo {title} {Hydrodynamics of lorentz symmetric systems: a quantum monte carlo study.}}\ } (\bibinfo {year} {2024}),\ \bibinfo {note} {{University of Würzburg, doi: \href{https://doi.org/10.58160/2EPKDQ7CTXTEBCP2}{10.58160/2EPKDQ7CTXTEBCP2}}}\BibitemShut {NoStop}%
\bibitem [{\citenamefont {{J\"{u}lich Supercomputing Centre}}(2019)}]{JUWELS}%
  \BibitemOpen
  \bibfield  {author} {\bibinfo {author} {\bibnamefont {{J\"{u}lich Supercomputing Centre}}},\ }\href {\doibase 10.17815/jlsrf-5-171} {\bibfield  {journal} {\bibinfo  {journal} {Journal of large-scale research facilities}\ }\textbf {\bibinfo {volume} {5}} (\bibinfo {year} {2019}),\ 10.17815/jlsrf-5-171}\BibitemShut {NoStop}%
\bibitem [{\citenamefont {M\"uller}\ \emph {et~al.}(2009{\natexlab{b}})\citenamefont {M\"uller}, \citenamefont {Schmalian},\ and\ \citenamefont {Fritz}}]{PhysRevLett.103.025301}%
  \BibitemOpen
  \bibfield  {author} {\bibinfo {author} {\bibfnamefont {M.}~\bibnamefont {M\"uller}}, \bibinfo {author} {\bibfnamefont {J.}~\bibnamefont {Schmalian}}, \ and\ \bibinfo {author} {\bibfnamefont {L.}~\bibnamefont {Fritz}},\ }\href {\doibase 10.1103/PhysRevLett.103.025301} {\bibfield  {journal} {\bibinfo  {journal} {Phys. Rev. Lett.}\ }\textbf {\bibinfo {volume} {103}},\ \bibinfo {pages} {025301} (\bibinfo {year} {2009}{\natexlab{b}})}\BibitemShut {NoStop}%
\bibitem [{\citenamefont {Buividovich}\ and\ \citenamefont {Polikarpov}(2012)}]{Buividovich12}%
  \BibitemOpen
  \bibfield  {author} {\bibinfo {author} {\bibfnamefont {P.~V.}\ \bibnamefont {Buividovich}}\ and\ \bibinfo {author} {\bibfnamefont {M.~I.}\ \bibnamefont {Polikarpov}},\ }\href {\doibase 10.1103/PhysRevB.86.245117} {\bibfield  {journal} {\bibinfo  {journal} {Phys. Rev. B}\ }\textbf {\bibinfo {volume} {86}},\ \bibinfo {pages} {245117} (\bibinfo {year} {2012})}\BibitemShut {NoStop}%
\bibitem [{\citenamefont {Tang}\ \emph {et~al.}(2024)\citenamefont {Tang}, \citenamefont {Yudhistira}, \citenamefont {Chattopadhyay}, \citenamefont {Ulybyshev}, \citenamefont {Sengupta}, \citenamefont {Assaad},\ and\ \citenamefont {Adam}}]{PhysRevB.110.155120}%
  \BibitemOpen
  \bibfield  {author} {\bibinfo {author} {\bibfnamefont {H.-K.}\ \bibnamefont {Tang}}, \bibinfo {author} {\bibfnamefont {I.}~\bibnamefont {Yudhistira}}, \bibinfo {author} {\bibfnamefont {U.}~\bibnamefont {Chattopadhyay}}, \bibinfo {author} {\bibfnamefont {M.}~\bibnamefont {Ulybyshev}}, \bibinfo {author} {\bibfnamefont {P.}~\bibnamefont {Sengupta}}, \bibinfo {author} {\bibfnamefont {F.~F.}\ \bibnamefont {Assaad}}, \ and\ \bibinfo {author} {\bibfnamefont {S.}~\bibnamefont {Adam}},\ }\href {\doibase 10.1103/PhysRevB.110.155120} {\bibfield  {journal} {\bibinfo  {journal} {Phys. Rev. B}\ }\textbf {\bibinfo {volume} {110}},\ \bibinfo {pages} {155120} (\bibinfo {year} {2024})}\BibitemShut {NoStop}%
\end{thebibliography}%

\end{document}